\documentstyle[12pt, epsf]{article}
\def\ZZZ{{\hbox{ Z\kern-1.6mm Z}}}
\newcommand{\beq}{\begin{equation}}
\newcommand{\eeq}{\end{equation}}
\newcommand{\bea}{\begin{eqnarray}}
\newcommand{\eea}{\end{eqnarray}}
\newcommand{\R}{\rangle}
\newcommand{\lt}{\left}
\newcommand{\rt}{\right}

\newcommand{\one}{{\hbox{ 1\kern-1.2mm l}}}

\newcommand{\T}{{\cal T}}
\newcommand{\g}{\gamma}
\newcommand{\gb}{\bar \gamma}

\newcommand{\mv}{M^{V}}
\newcommand{\ms}{M^{S}}
\newcommand{\mc}{M^{C}}
\newcommand{\mbv}{\bar M^{V}}
\newcommand{\mbs}{\bar M^{S}}
\newcommand{\mbc}{\bar M^{C}}
\newcommand{\bps}{\hbox{BPS}}
\newcommand{\nbps}{\mathbin{\hbox{BPS}\mkern-20mu\Big /}}
\newcommand{\al}{\alpha}
\newcommand{\at}{\tilde \alpha}
\newcommand{\ev}{\hbox{even}}
\newcommand{\odd}{\hbox{odd}}

\newcommand{\nNp}{\bar{\cal N}_p}
\newcommand{\nTp}{\bar{\cal T}_p}

\newcommand{\Ft}{\tilde F}
\newcommand{\St}{\tilde S}
\newcommand{\psit}{\tilde \psi}
\newcommand{\zb}{\bar z}
\newcommand{\wb}{\bar w}
\newcommand{\Jt}{\tilde J}

\newcommand{\CS}{{\cal S}}
\newcommand{\ze}{\zeta}
\newcommand{\zeb}{\bar \zeta}
\newcommand{\dt}{\delta}

\newcommand{\D}{\Delta}
\newcommand{\p}{\prime}
\newcommand{\eps}{\epsilon}
\newcommand{\M}{{\cal M}}
\newcommand{\delb}{\bar \delta}
\newcommand{\Del}{\Delta}
\newcommand{\Delb}{\bar \Delta}
\newcommand{\lam}{\lambda}
\newcommand{\lamb}{\bar \lambda}
\newcommand{\K}{{\cal K}}
\newcommand{\KD}{{\cal K_D}}

\newcommand{\sectiono}[1]{\section{#1}\setcounter{equation}{0}}

\begin{document}
{}~
{}~
\hfill\vbox{\hbox{UK/04-24} \hbox{hep-th/0411103}}\break

\vskip .6cm

\centerline{\Large \bf Non-BPS D-Branes in}
\medskip
\centerline{\Large \bf Light-Cone Green-Schwarz Formalism}

\medskip

\vspace*{4.0ex}

\centerline{\large \rm Partha Mukhopadhyay }

\vspace*{4.0ex}

\centerline{\large \it  Department of Physics and Astronomy}

\centerline{\large \it  University of Kentucky, Lexington, KY-40506, U.S.A.}

\medskip

\centerline{E-mail: partha@pa.uky.edu}

\vspace*{5.0ex}

\centerline{\bf Abstract} \bigskip

Non-BPS D-branes are difficult to describe covariantly in a manifestly
supersymmetric formalism. For definiteness we concentrate on
type IIB string theory in flat background in light-cone
Green-Schwarz formalism. We study both the boundary state and
the boundary conformal field theory descriptions of these D-branes
with manifest $SO(8)$ covariance
and go through various consistency checks. We analyze Sen's original
construction of non-BPS D-branes given in terms of an orbifold
boundary conformal field theory. We also directly study the relevant
world-sheet theory by deriving the open string boundary condition
from the covariant boundary state. Both these methods give the same
open string spectrum which is consistent with the boundary state, as
required by the world-sheet duality. The boundary condition found in
the second method is given in terms of bi-local fields that are
quadratic in Green-Schwarz fermions. We design a special ``doubling
trick'' suitable to handle such boundary conditions and prescribe
rules for computing all possible correlation functions without
boundary insertions. This prescription has been tested by computing
disk one-point functions of several classes of closed string states
and comparing the results with the boundary state computation.

\vfill \eject

\tableofcontents

\baselineskip=18pt

\sectiono{Introduction and Summary} \label{s:intro}

Although the Neveu-Schwarz-Ramond (NSR) formalism \cite{polchinski98}
is very advantageous for describing string theory in NS-NS
backgrounds, it suffers from several drawbacks.
These include lack of manifest space-time supersymmetry and
difficulty in quantizing strings in RR backgrounds. Recent progress in
Berkovits' pure spinor formalism \cite{berkovits} is encouraging in
the direction of finding a tool free of such problems\footnote{See
\cite{others} for various extensions of the pure spinor formalism and 
other relevant works.}.
It is also desirable that we be able to describe all the valid objects
in string theory in whatever language we find the most convenient to work
with. Non-BPS D-branes are certain non-supersymmetric non-perturbative
solutions in string theory that are usually described in NSR
formalism (see, for example, 
\cite{sen9803, sen9805, bergman98, sen9809, frau99, senrev, lerda99}).
Unlike the BPS D-branes \cite{polchinski96, green96} they
are difficult to describe in a manifestly supersymmetric formalism
where the basic world-sheet fields are space-time
spinors \cite{mukhopadhyay00}. In this paper we return to the
same problem discussed in \cite{mukhopadhyay00}, namely:
how to describe non-BPS D-branes in type II string theories in flat
space using light-cone Green-Schwarz (GS) formalism? Although this is
not a covariant description \cite{green87}, it captures the basic
problem of describing a non-BPS D-brane in a manifestly supersymmetric
setup. A particular motivation to continue the above study is to
look for the analogous D-branes in type IIB pp-wave background with
RR-flux where
the string quantization has been successfully done only in light-cone GS
formalism \cite{metsaev01}. This background is particularly interesting
for exploring such questions in the context of AdS/CFT
duality \cite{berenstein02, maldacena03}.

In NSR formalism the world-sheet fermions are space-time vectors and
satisfy simple linear boundary conditions for both the BPS and non-BPS
D-branes. This is not true for non-BPS D-branes in GS formalism where
the world-sheet fermions are also space-time fermions\footnote{The
problem lies only in the fermionic sector of the world-sheet. Unless
it is explicitly mentioned, the bosonic part will not play any role in
any of our discussions.}. In spite of this difficulty non-BPS D-branes in
flat background  can be defined in a current algebraic framework
\cite{mukhopadhyay00} where the type II string theories in light-cone
gauge are considered to be particular realizations of $SO(8)$
Kac-Moody algebra at level $k=1$ \cite{goddard86}. Boundary condition 
on the local currents relevant to either a BPS or a non-BPS D-brane
can be easily written down in terms of an automorphism which is simply
the collection of reflections along the Neumann directions of the brane.
This implies that for even dimensional world-volume this automorphism
is ``inner'' whereas for odd dimensional world-volume this is
``outer''. It is the particular property of the fermionic variables
used in GS formalism that it is difficult to describe ``inner
automorphism boundary conditions'' in the realization of type IIA and 
``outer automorphism boundary conditions'' in the realization of type 
IIB. These precisely correspond to the non-BPS D-branes in the two theories. 

In current algebraic framework a non-BPS D-brane is given by a 
particular Ishibashi state. This is simply a sum over the left-right 
symmetric basis states in a particular highest weight representation 
with the outer automorphism (for type IIB) concerned acted on the
right part \cite{ishibashi88, kato96}. One way to
see the problem with the type IIB non-BPS D-branes is 
that the action of the outer automorphism is not well defined on the 
right moving GS fermions, rather it is well
defined only on the states spanning the particular highest weight
representation involved \cite{mukhopadhyay00}. Therefore to get an 
organized expression of
the Ishibashi state in terms of the GS fermions a choice of the
basis has to be made which spans only the relevant subspace. In 
\cite{mukhopadhyay00} this basis was constructed
explicitly in terms of the current modes. The problem with this choice
is that the free module of the current modes contains null states
which have to be removed by hand, a procedure which destroys manifest 
$SO(8)$ covariance. Indeed the analysis in \cite{mukhopadhyay00} for 
the boundary states corresponding to branes with various orientations 
looks cumbersome. One has the similar problem also with the type IIA 
non-BPS D-branes.

Concentrating on the type IIB theory for definiteness we arrive at the
covariant expression for a non-BPS boundary state 
by manipulating the Green-Schwarz basis states spanning the relevant
subspace in a particular way. Similar expression was first extracted, 
using a nice trick, from the NSNS part of a BPS boundary state in
\cite{nemani} (see section \ref{s:discussion} for comments on this
method). Our result is very similar to but not exactly same as that
found in \cite{nemani}. This covariant expression appears to be much
more complicated than that of a BPS D-brane \cite{green96}. However, 
it turns out that one can
use a suitable bosonization and refermionization technique to
construct a new anti-holomorphic spinor field $\bar S^a(\zb)$ out of the 
right-moving GS fermion $\tilde S^a(\zb)$ so that the non-BPS boundary 
state, when expressed in terms of the new oscillators
$\bar S^a_n$, looks exactly like the NSNS part of a BPS D-instanton 
(zero world-volume dimension) boundary state in type IIB theory. There
are certain computations for which this new form of the boundary state
turns out to be very helpful. For example, as discussed in appendix
\ref{a:bc}, one can derive the open string boundary condition from this 
form of the boundary state quite easily.

We also study the open string description of the system given in terms
of a boundary conformal field theory (BCFT) and go through various
consistency checks of the whole discussion of boundary state and BCFT.
We partially discuss two approaches to the BCFT problem, both 
producing the same
result for the open string spectrum\footnote{This spectrum is already
known from the study in NSR formalism. Here we get the Green-Schwarz
organization of the open string states.}. This contains an R and an 
NS sector states. The R sector states have Bose-Fermi degeneracy at 
every level while the NS sector states, which include the tachyon, do not.
We use this spectrum and the covariant boundary state to check the 
open-closed duality on the world-sheet. 

In the first approach we analyze the original method due to Sen 
\cite{senrev} in GS formalism. In this
method type IIB (IIA) string theory is regarded as type IIA (IIB)
string theory with an orbifold projection by $(-1)^{\tilde F}$, 
where $\tilde F$
is the space-time fermion number contributed by the anti-holomorphic 
part of the world-sheet theory. To construct a non-BPS D-brane in type
IIB one starts out with a suitable pair of BPS D-brane and
anti-D-brane in type IIA and then projects out the whole configuration
by $(-1)^{\tilde F}$. Since we are dealing with an orbifold BCFT and
we know how to deal with the parent theory which involves only BPS
D-branes one does not expect to see any obvious problem in
computing the correlation functions. But we have not gone through the
explicit analysis. 

In the second approach, which is more direct, a non-BPS D-brane 
in the type IIB theory 
is described by the standard type IIB GS action with a certain
boundary condition on the world-sheet fermions. Traditionally, the
boundary condition corresponding to a D-brane is obtained by first
varying the classical world-sheet action (with boundary) with respect
to the basic fields and then setting the boundary term to
zero. In the present context this method gives all the BPS boundary
conditions quite easily, but it is hard to guess what a non-BPS
boundary condition should be. However, the covariant form of the non-BPS
boundary state and the reduction of its form to that of the NSNS part
of a BPS D-instanton boundary state (as described above) enable us to 
derive the required open string boundary condition. A BPS boundary
state satisfies a certain linear gluing condition relating the left
and right moving oscillators \cite{green96}. From this it is
easy to derive the quadratic gluing condition satisfied by the NSNS
part of this boundary state\cite{nemani}. This tells us what gluing
condition a non-BPS boundary state should satisfy in terms of the
$S^a_n$ and $\bar S^a_n$ oscillators. Converting this gluing condition
into the open string channel and using the bosonization and
refermionization trick in the reverse direction we finally end up
deriving the open string boundary condition in terms of the GS
fermions $S^a(z)$ and $\tilde S^a(\zb)$. This boundary condition
relates certain left and right moving fields which are bi-local and 
quadratic in GS fermions. Having obtained this boundary condition one
can explain how that is compatible with the condition that one obtains 
by setting the boundary variation to zero. 

It would be nice to be able to quantize the world-sheet theory in the
``direct approach'' described above. We have
not tried to do this directly with such a bi-local boundary condition. 
Rather we have taken an approach where one comes up with a minimal set
of rules for computing all possible correlation functions of this
boundary theory. In this paper we suggest that this is possible at
least for the correlation functions without boundary insertions by
explicitly prescribing the relevant rules. This involves extending the
bi-local boundary conditions to all possible spin fields in the theory
and designing a special ``doubling trick'' for handling such boundary
conditions. Using our prescription we have computed disk one-point
functions for several classes of closed string states with the
puncture at the center of the disk. It has been shown that the results
match, quite non-trivially, with the computation done with the
covariant boundary state. The particular classes of closed string
states have been chosen so as to verify certain numerical factors
explicitly appearing in the covariant boundary state.
The open string spectrum found in the ``orbifold approach'' discussed
first can also be derived quite easily using the above doubling
trick.

We have also made some naive attempts to look for the analogous
D-branes in type IIB pp-wave background. In absence of a current
algebraic framework we have tried to exploit the
algebraic similarity between the string quantization in flat and
pp-wave backgrounds. The algebraic structure of the zero modes is
quite different in these two backgrounds and the attempts fail
precisely because of some subtlety involving the fermionic zero modes.

The rest of the paper is organized as follows. Sec.\ref{s:BS} contains
the boundary state description. We first summarize the final result
for the covariant boundary state, then discuss the bosonization and
refermionization procedure to arrive at the simpler form. We discuss the
BCFT description in sec.\ref{s:BCFT} where the two approaches are
presented in the two subsections \ref{ss:orbifold} and
\ref{ss:direct}. Some relevant points including some future directions
are discussed in sec.\ref{s:discussion}. Various technical details
have been presented in several appendices.

\sectiono{Boundary State Description}
\label{s:BS}

For definiteness we shall consider type IIB string theory. 
All the D-branes that we shall consider here are instantonic in the
sense that both the light-cone coordinates ($X^0\pm X^9$) satisfy 
Dirichlet boundary condition \cite{green94, green96}. The 
$(p+1)$-dimensional ($-1\leq p\leq 7$) world-volume of
a D$p$-brane is therefore euclidean. Without any loss of generality 
we shall always consider a D$p$-brane (either BPS or non-BPS) to be
aligned along coordinate directions such as 
$(x^{I_1}, \cdots , x^{I_{p+1}})$ for notational simplicity.
Odd and even values of $p$ correspond to BPS and non-BPS D-branes
respectively in type IIB. In each case we define a set of matrices 
giving the vector, spinor and conjugate spinor
representations\footnote{Recall that because of the triality symmetry
\cite{green87} of the associated Dynkin diagram there are three lowest
($8$) dimensional representations of $SO(8)$.}
of the set of reflections along the Neumann directions. For $p$ odd, which
corresponds to a BPS D-brane, we define,
\bea
\mv_{IJ} &=& \lam_I \delta_{IJ}~,\quad 
\lam_I =  \lt \{ \begin{array}{cl}
1 & \hbox{if $x^I$ is a Dirichlet direction}~, \\
-1 & \hbox{if $x^I$ is a Neumann direction}~, 
\end{array} \rt. \cr
\ms_{ab} &=& \lt( \gamma^{I_1} \bar{\gamma}^{I_2}\gamma^{I_3} 
\bar{\gamma}^{I_4}\cdots \gamma^{I_p}\bar{\gamma}^{I_{p+1}}\rt)_{ab}~, 
\cr
\mc_{\dot a\dot b} &=& \lt( \bar{\gamma}^{I_1} \gamma^{I_2}
\bar{\gamma}^{I_3} \gamma^{I_4}\cdots \bar{\gamma}^{I_p}
\gamma^{I_{p+1}} \rt)_{\dot a\dot b}~,
\label{MVMSMC}
\eea
where the gamma matrices in the last two equations are placed with a
chosen ordering of the Neumann indices. These expressions correspond
to the following block off diagonal form of the $16$-dimensional gamma 
matrices:
$\displaystyle{ \Gamma^I = \pmatrix{0 & \gamma^I_{a\dot b} \cr
\bar \gamma^I_{\dot a b} & 0 }}$.
Unless it is explicitly mentioned otherwise we shall always consider a
real representation (as in \cite{green87}), in which case $\gb^I =
(\g^I)^{\rm T}$. 
For $p$ even, corresponding to a non-BPS 
D-brane, we denote these matrices by $\mbv_{IJ}$, $\mbs_{a\dot b}$ and
$\mbc_{\dot a b}$ respectively and use the similar definitions. All
the $M$ matrices defined this way are orthogonal. The fermionic gluing
condition satisfied by the boundary state of a BPS D-brane is simply 
given by \cite{green96},
\beq
(S^a_n + i \eta \ms_{ab} \tilde S^b_{-n}) |\bps, p, \eta \R = 0~,
\quad \forall n \in \ZZZ. 
\label{bpsgluing}
\eeq
But notice that because of the particular index structure of 
$\mbs_{a\dot b}$ the boundary state for a non-BPS D-brane does not
satisfy such a simple gluing condition. Although the above gluing
condition does not have a simple generalization to the non-BPS case it
was shown in \cite{mukhopadhyay00} that the current algebraic
construction of the BPS boundary states does have. This method has
been reviewed in appendix \ref{a:CAconstruction}. It turns out that
a covariant expression for the non-BPS boundary state can be obtained
by manipulating the Green-Schwarz basis expansion in an Ishibashi
state in a certain way (see appendix \ref{a:CAderivation}). 
Similar expression was first extracted, using a nice trick, from the 
NSNS part of a BPS boundary state in \cite{nemani}. The result
reported here is very similar to but not
exactly same as that found in \cite{nemani}. Here we shall summarize 
the final result. The boundary state of a non-BPS D-brane in type-IIB 
situated at the origin of the transverse directions is given by,
\bea
|\nbps~ \R = \nNp \int dk_{\perp} \exp \lt(\sum_{n=1}^{\infty}
{1\over n} \al^I_{-n} \mbv_{IJ} \at^J_{-n} \rt) |e\R \R \otimes 
|k_{\perp}\R~,
\label{nonbps}
\eea
where $\nNp$ is a normalization constant\footnote{We shall show in
appendix \ref{a:duality} that the boundary state (\ref{nonbps})
satisfies the open-closed world-sheet duality for the expected value
of $\nNp$.}. $k_{\perp}$ is the momentum in the transverse space which
includes $x^0$ and $x^9$. As explained in appendix
\ref{a:CAderivation}, $|e\R\R$ is a current algebra Ishibashi state.
Its covariant expression is given by, with some additional
prescription that we shall spell out shortly,
\bea
|e\R \R &=& | \mbv, \ev \R + |\mbv, \odd \R ~, \cr
|\mbv \ev \R &=& \cosh \lt(\sqrt{X_{\mbv}}\rt) \mbv_{IJ}
|I\R \otimes \widetilde{|J\R} ~, \cr
|\mbv, \odd \R &=& - {\sinh \lt(\sqrt{X_{\mbv}}\rt) \over 
\sqrt{X_{\mbv}} }
Y^{\dot a \dot b}_{\mbv} |\dot a\R \otimes \widetilde{|\dot b\R} ~.
\label{nonbpstwo}
\eea
The operators $X_{\mbv}$ and $Y^{\dot a \dot b}_{\mbv}$ are given by,
\bea
X_{\mbv} &=& \sum_{m,n>0} \lt[{1\over 8} J_{-m,-n} \tilde J_{-m,-n} +
{1 \over 16} \sum_{IJ} \lamb_{\{IJ\}} 
J^{IJ}_{-m,-n} \tilde J^{IJ}_{-m,-n} + \rt. \cr 
&& \lt. 
{2\over 384} 
\sum_{\{IJKL\} \in \K} \lamb_{\{IJKL\}} J^{IJKL}_{-m,-n} \tilde 
J^{IJKL}_{-m,-n}\rt]~,
\label{XM}
\eea
\bea
Y^{\dot a \dot b}_{\mbv} &=& \sum_{n>0} \lt[ {1\over 8} 
\sum_I \lamb_I \g^I_{a\dot a} \g^I_{b\dot b} S^a_{-n}  
\tilde S^b_{-n} + 
{1\over 48} \sum_{IJK} \lamb_{\{IJK\}} 
\g^{IJK}_{a\dot a} \g^{IJK}_{b\dot b} S^a_{-n}  
\tilde S^b_{-n} \rt]~,
\label{YM}
\eea
where we have denoted the eigenvalues of $\mbv$ by $\lamb_I$ and used
the notation: $\lamb_{\{IJ\cdots\}} = \lamb_I \lamb_J \cdots$. 
We also used the following definitions \cite{nemani},
\bea
J_{mn}=S^a_m S^a_n, \quad J^{IJ}_{mn} = \g^{IJ}_{ab}S^a_m S^b_{n}~,
\quad J^{IJKL}_{mn} = \g^{IJKL}_{ab}S^a_m S^b_{n} ~,
\label{Jmn}
\eea
and similarly for the the right moving variables. A multi-indexed
gamma matrix is antisymmetric under interchange of any of the two
indices and it is defined in the following way,
\bea
\gamma^{I_1I_2 \cdots I_n} &=& \lt\{\begin{array}{ll} 
\g^{I_1} \bar \g^{I_2} \cdots ~, & I_1 \neq I_2 \neq \cdots \neq I_n ~,\\
0~, & \hbox{otherwise} \end{array} ~.\rt.  
\label{gammas}
\eea
Although the last two equations in (\ref{nonbpstwo}) have been given 
compact forms in terms of the $\cosh$ and $\sinh$ operators they are 
actually understood as series expansions where only integer powers of 
$X_{\mbv}$ appear. The overall sign of the state $|\mbv, \odd\R$ in
(\ref{nonbpstwo}) depends on a certain convention while that of the
state $|\mbv,\ev\R$ does not. Our convention is to take $\St^a(\zb)$ to
commute with the left moving vector spin field $\psi^I(z)$ and anti-commute  
with the conjugate spinor spin field $S^{\dot a}(z)$\footnote{This
  convention is just opposite to that followed in
  \cite{mukhopadhyay00}. \label{convention}}. 
Finally, let us explain the last term on
the right hand side of eq.(\ref{XM}). We divide the set of all
possible sets of four indices $\{I,J,K,L\}$ into two subsets of equal 
order: $\K$ and $\KD$ such that for every element $\{I,J,K,L\}$ in
$\K$ there is a dual set of indices $\{M,N,O,P\}$ in $\KD$ satisfying
$\eps^{IJKLMNOP} \neq 0$. Since $J^{IJKL}_{mn}$ are self dual
operators,
\bea
J^{IJKL}_{mn}= {1\over 4!} \eps^{IJKLMNOP}J^{MNOP}_{mn}~,
\label{self-dualJ}
\eea
the number of independent components is same as the order of $\K$ or
$\KD$ modulo the anti-symmetrization among the four indices. The
summation in the last term on the right hand side of eq.(\ref{XM}) is
restricted to the elements of $\K$. Notice that $\K$ is not unique,
it has to be chosen by hand and the state $|e\R\R$ does depend on
this choice. We would like to point out that this choice is at the
very level of basis construction in GS formalism. The GS variables
corresponding to different choices are related in a complicated
manner through bosonization and refermionization\footnote{I would like
to thank A. Sen for emphasizing this point.}. To see this more clearly
and to spell out the additional implicit prescription in 
eq.(\ref{nonbpstwo}) through an example, let us concentrate on the
states at level 3. This is the minimum level where the operators 
$J^{IJKL}_{m,n}$ appear in the basis construction. In NSR formalism
the rank 5 tensor states at this level are given by 
$\psi^I_{-1/2}\psi^J_{-1/2}\psi^K_{-1/2}\psi^L_{-1/2}\psi^M_{-3/2}
|0\R_{NS}$. There is no self duality condition on the first
four indices and there are 560 independent components. In GS formalism
all these states have to come from two sets of states:
$\g^{IJKL}_{ab}S^a_{-1}S^b_{-2}|M\R$ and $\g^{IJ}_{ab}\g^{KLM}_{c\dot d}
S^a_{-1}S^b_{-1}S^c_{-1}|\dot d\R$, each contributing 280 independent
states\footnote{The tensor index structure of 
$\g^{IJ}_{ab}\g^{KLM}_{c\dot d} S^a_{-1}S^b_{-1}S^c_{-1}|\dot d\R$
does not quite lead to 280 independent states. The counting is more
complicated because of the antisymmetry among all the three
oscillators. But there is an indirect way: These states, along with
the state $\g^{IJ}_{ab}\g^K_{c\dot d} S^a_{-1}S^b_{-1}S^c_{-1}|\dot
d\R $ make up all the states in $ S^a_{-1}S^b_{-1}S^c_{-1}|\dot d\R$
which can be counted to have 448 independent states. One can count the
number of independent rank 3 tensor states from NSR formalism without
any trouble. This also turns out to be 448. In addition to 
$\g^{IJ}_{ab}\g^K_{c\dot d} S^a_{-1}S^b_{-1}S^c_{-1}|\dot d\R $ they
also include $\g^{IJ}_{ab}S^a_{-1}S^b_{-2}|K\R$ and 
$\g^{IJK}_{a\dot b}S^a_{-3}|\dot b\R$ whose counting (in GS formalism)
gives 224 and 56 respectively. This gives (448-224-56)=168 for
$\g^{IJ}_{ab}\g^K_{c\dot d} S^a_{-1}S^b_{-1}S^c_{-1}|\dot d\R $. 
Therefore we need 280 for 
$\g^{IJ}_{ab}\g^{KLM}_{c\dot d} S^a_{-1}S^b_{-1}S^c_{-1}|\dot d\R$ to
make up 448 for $ S^a_{-1}S^b_{-1}S^c_{-1}|\dot d\R$. } making up the
total of 560. Therefore to specify the basis one has to choose 280
states from each of the two groups. This choice, for the group of
states $\g^{IJKL}_{ab}S^a_{-1}S^b_{-2}|M\R$ , is given by $\K$. States
for which $\{I,J,K,L\}\in \K_D$ should come from the other group: 
$\g^{IJ}_{ab}\g^{KLM}_{c\dot d} S^a_{-1}S^b_{-1}S^c_{-1}|\dot d\R$.
Notice that the two groups of states appear in $|\mbv, \ev\R$ and
$|\mbv, \odd\R$ respectively in eq.(\ref{nonbpstwo}). Reduction of
the first group of states to the independent ones (modulo 
anti-symmetrization) as mentioned above is taken care of explicitly 
by the constrained sum in the last term in eq.(\ref{XM}). As can be
seen easily by inspection that this is not true for the second group
of states. This reduction has to be done by hand in the proper terms. 
Once it is done, one ends up having the rank 5 tensor states, with the
same properties of the NSR states mentioned above, with free sum over
all the indices in the expansion of the boundary state. Once this 
procedure is followed at all the levels, the boundary state, when expanded
in terms of the basis states, does not depend on $\K$. This is
guaranteed, as the boundary state written in this fashion should take
the same form as that obtained in light-cone NSR formalism (though it
is difficult to prove this in GS formalism). It is only when the basis
states are written in terms of the GS variables, the $\K$ dependence
comes in.

\vspace{15pt}
\centerline{\it Bosonization and Refermionization}
\vspace{10pt}

\noindent
We shall now show that the complicated expression of $|e\R\R$ in
eq.(\ref{nonbpstwo}) can be given a much simpler form in terms of
suitably defined variables which are obtained by bosonizing and
refermionizing (say) the antiholomorphic triad 
$\lt\{ \St^a(\zb), \St^{\dot a}(\zb), \tilde \psi^I(\zb)\rt\}$ on the
full plane in a certain manner. Let us first define a new spinor in
the following way,
\bea
\hat S^{\dot a}(\zb) = \mbc_{\dot a a} \tilde S^a(\zb)~.
\label{Shat}
\eea
Using the following relations that can be derived from the
definitions of the $\bar M$ matrices given below eq.(\ref{MVMSMC}),
\bea
\lt(\mbs \rt)^{\rm T} &=& (-1)^{p/2} \mbc ~,  \quad (\hbox{recall
$p$ is even})~, \cr
\mbc \g^I \lt( \mbs \rt)^{\rm T} &=& \lamb_I \bar
\g^I ~, 
\label{relationtwo}
\eea
one can show,
\bea
\hat S^{\dot a}(\zb) \gb^{IJ\cdots}_{\dot a\dot b} 
\hat S^{\dot b}(\wb) =
\lamb_{\{IJ\cdots \}} \St^a(\zb) \g^{IJ\cdots}_{ab} \St^b(\wb) ~,
\label{Shat-Stld}
\eea
where we mean to have a product of even number of gamma matrices as
indicated by the index structure. The multi-indexed $\bar \g$ matrices
are defined by replacing $\g$'s by $\bar \g$'s and vice versa in 
eq.(\ref{gammas}),
\bea
\bar \gamma^{I_1I_2 \cdots I_n} = \lt\{\begin{array}{ll} \bar \g^{I_1} 
\g^{I_2} \cdots ~, & \quad I_1 \neq I_2 \neq \cdots \neq I_n ~,\\
0~, &\quad \hbox{otherwise}~. \end{array} \rt. 
\label{gammabars}
\eea
Notice that although 
$\hat S^{\dot a} \gb^{IJKL}_{\dot a\dot b} \hat S^{\dot b}$ is an 
anti-self-dual operator (as can be easily verified by using the last
equation in (\ref{traces})) while
$\St^a(\zb) \g^{IJKL}_{ab} \St^b(\wb)$ is self-dual,
eq.(\ref{Shat-Stld}) holds because of the following relation,
\bea
\lamb_{\{IJKL\}}=-\lamb_{\{MNOP\}}~, \quad \hbox{whenever } 
\eps^{IJKLMNOP} \neq 0~.
\label{lamb-duality}
\eea
We now suitably bosonize and refermionize $\hat S^{\dot
a}(\zb)$ to obtain a new fermion $\bar S^a(\zb)$ such that,
\bea
\bar S^{a}(\zb) \bar S^{a}(\wb) &=& 
\hat S^{\dot a}(\zb) \hat S^{\dot a}(\wb)~, \cr
\bar S^{a}(\zb) \g^{IJ}_{a b} 
\bar S^{b}(\wb) &=& \hat S^{\dot a}(\zb) \gb^{IJ}_{\dot a\dot b} 
\hat S^{\dot b}(\wb)~,\cr
\bar S^{a}(\zb) \g^{IJKL}_{a b} 
\bar S^{b}(\wb) &=& \lt\{
\begin{array}{ll}
\hat S^{\dot a}(\zb) \gb^{IJKL}_{\dot a\dot b} \hat S^{\dot b}(\wb) & 
\hbox{for } \{I,J,K,L\} \in \K ~, \\
-\hat S^{\dot a}(\zb) \gb^{IJKL}_{\dot a\dot b} \hat S^{\dot b}(\wb) &
\hbox{for } \{I,J,K,L\} \in \KD ~.
\end{array}  \rt.
\label{Sbar-Shat}
\eea
Denoting the spin fields of $\bar S^a(\zb)$ by $\bar \psi^I(\zb)$ and
$\bar S^{\dot a}(\zb)$ one can also relate them to the spin fields of 
$\St^a(\zb)$,
\bea
\bar \psi^I(\zb) &=& \lamb_I \tilde \psi^I(\zb)~, \label{BR1} \cr
\bar S^{\dot a}(\zb)\gb^{IJ\cdots}_{\dot a\dot b}\bar S^{\dot b}(\wb)
&=& \lamb_{\{IJ\cdots \}}
\St^{\dot a}(\zb)\gb^{IJ\cdots}_{\dot a \dot b}\St^{\dot b}(\bar w)~, \cr
\bar S^a(\zb) \g^{IJ\cdots}_{a\dot a} \bar S^{\dot a}(\wb) &=&
\lamb_{\{IJ\cdots \}} \St^a(\zb) \g^{IJK}_{a \dot a} \St^{\dot a}(\bar w)~, 
\label{spinbar-spintld}
\eea
where the second and third equations have products of even and odd
number of gamma matrices respectively as indicated by the index structures. 
In case of four vector indices $(I,J,K,L)$ the second equation is
understood to be true for $\{I,J,K,L\}\in \K$. For $\{I,J,K,L\}\in
\KD$ the equation will have an additional factor of $(-1)$.
Equations
(\ref{Shat-Stld}), (\ref{Sbar-Shat}) and (\ref{spinbar-spintld})
clearly show that $\lt\{ \bar S^a(\zb), \bar S^{\dot a}(\zb), 
\bar \psi^I(\zb)\rt\}$ corresponds to the same gamma matrix
representation in the frame transformed by $\mbv$ as that
corresponding to 
$\lt\{ \St^a(\zb), \St^{\dot a}(\zb), \tilde \psi^I(\zb)\rt\}$ in the
untransformed frame. Therefore, for the zero modes $\bar S^a_0$, one
has,
\bea
\bar S^a_0 \bar{|I\R} = {1\over \sqrt{2}} \g^I_{a\dot a} \bar{|\dot
a\R}~, 
\label{Sbar0}
\eea
where $\bar{|I\R}$ and $\bar{|\dot a\R}$ are the ground states created
by the spin fields $\bar \psi^I(\zb)$ and $ \bar S^{\dot a}(\zb)$
respectively.

From the above discussion, it is not difficult to convince oneself
that the NS-sector states (following to the nomenclature of NSR 
formalism\footnote{According to the notation in eq.\ref{spectIIA}, these
states span the irreducible representation
$\Pi^{(e)}$.}) on the anti-holomorphic side written in
terms of the oscillators $\St^a_n$'s and the states 
$\widetilde{|I\R}$ and $\widetilde{|\dot a\R}$ can be translated,
without much trouble, in terms of the oscillators $\bar S^a_n$'s and the
states $\bar{|I\R}$ and $\bar{|\dot a\R}$. This allows us to rewrite
the state $|e\R\R$ in eq.(\ref{nonbpstwo}) in terms of these new variables.
In fact, using eq.(\ref{lamb-duality}) it is not difficult to see that
the translated expression
should simply take the same form as in eq.(\ref{nonbpstwo}) with
$\mbv$ set to identity. One then uses the algebraic manipulation of 
appendix \ref{a:CAderivation} in the reverse direction to finally 
arrive at the following expression,
\bea
|e\R\R = {1\over 2} \lt(|\hbox{BPS}, -1, + \R_{\bar S} + |\hbox{BPS},
-1, -\R_{\bar S} \rt)~,
\label{e-nsns}
\eea
where $|\hbox{BPS}, -1,\eta \R_{\bar S}$ ($\eta =\pm 1$) is obtained,
up to a normalization constant, by replacing: $\tilde S^a_{-n} \to
\bar S^a_{-n}$, $\widetilde{|I\R} \to \bar{|I\R}$ and 
$\widetilde{|\dot a\R} \to \bar{|\dot a\R}$ in the fermionic part of a
BPS D-instanton (zero world-volume dimension) boundary state.
\bea
|\hbox{BPS}, -1,\eta \R_{\bar S} = 
\exp \lt(- i\eta \sum_{n>0,a} S^a_{-n} \bar S^a_{-n}\rt) \lt[\sum_I |I\R
\otimes \bar{|I\R} - i \eta \sum_{\dot a} |\dot a\R \otimes \bar{|\dot
a\R}\rt]~,
\label{Dinst}
\eea
It is interesting to notice that $|e\R\R$ in eq.(\ref{e-nsns}) takes
the form of the NSNS part of the BPS D-instanton boundary
state. Certain analysis, such as those in appendices \ref{a:duality} and
\ref{a:bc}, can be enormously simplified with this form.

We shall test the covariant form in (\ref{nonbpstwo}) by checking the
open-closed duality on the cylinder. We shall also explicitly verify
the numerical factors $1/8$, $1/16$, $2/384$ and $1/48$ appearing in
eqs.(\ref{XM}, \ref{YM}) by computing one-point functions of the
relevant classes of closed string states in a boundary conformal field
theory description. 
\sectiono{Boundary Conformal Field Theory Description}
\label{s:BCFT}
In this section we shall study the problem from the open string point
of view. We have discussed two approaches. The first one 
(subsec.\ref{ss:orbifold}) contains the analysis of Sen's original 
orbifold method in GS formalism. In the second approach 
(subsec.\ref{ss:direct}) we study the problem more directly in terms
of a BCFT with certain boundary condition. Both the methods produce
the same open string spectrum which is consistent, according to the
world-sheet duality, with the covariant boundary state
(appendix \ref{a:duality}). 

\subsection{Orbifold Approach}
\label{ss:orbifold}

Type IIB (IIA) superstring theory can be regarded as the $(-1)^{\Ft}$
orbifold of type IIA (IIB) theory. Here $\Ft$ is the space-time
fermion number contributed by the  right moving sector of the
world-sheet. The original construction of type IIB (IIA) non-BPS
D-branes due to Sen went through as follows: Take a non-supersymmetric
configuration of pair of coincident BPS D-brane and anti-D-brane in
type IIA (IIB) and then project the
configuration out by $(-1)^{\Ft}$. This gives the corresponding
non-BPS D-brane in type IIB (IIA). Here we shall outline the same
procedure in GS formalism. As an outcome of this analysis we find the
Green-Schwarz organization of the open string spectrum on a non-BPS
D-brane. The same result is also found in the second approach in
subsec.\ref{ss:direct}.
\vskip 10pt
\centerline{\it Type IIB Theory as $(-1)^{\tilde F}$ orbifold of Type IIA Theory}
\vskip 10pt
\noindent
Since the bosonic part is same in both the two theories and the
projection does not act on this we shall focus only on the fermionic
part of the theory. In light-cone gauge, this part of type IIA theory
can be viewed as the following particular realization of 
$\widehat{SO}(8)_{k=1}$\cite{goddard86, mukhopadhyay00}:
\bea
\hbox{type IIA} : \lt( \Pi^{(e)} \oplus \Pi^{(\bar \delta)} \rt)
\otimes \lt( \Pi^{(e)} \oplus \Pi^{(\delta)} \rt)~,
\label{spectIIA}
\eea
where $\Pi^{(w)}$ is the irreducible representation corresponding to
the highest weight $w$. $e$, $\delta$ and $\bar \delta$ are the
$SO(8)$ fundamental weights corresponding to the vector, spinor and
conjugate spinor representations respectively. In terms of the GS 
oscillators a generic state in type IIA, therefore, takes the
following form,
\bea
S^{a_1}_{-n_1} S^{a_2}_{-n_2} \cdots S^{a_N}_{-n_N}
\tilde S^{\dot a_1}_{-m_1} \tilde S^{\dot a_2}_{-m_2} \cdots 
\tilde S^{\dot a_{M}}_{-m_M} \pmatrix{|I\R \cr |\dot a\R} \otimes
\pmatrix{\widetilde{|J\R} \cr \widetilde{|a \R}}~.
\label{IIAstate}
\eea
Obviously, one needs to impose the left-right level-matching condition
on it. While $(-1)^{\Ft}$ acts as an identity operator on the
left-moving variables, its action on the right-moving variables is 
given by,
\bea
(-1)^{\Ft}:
\quad \tilde S^{\dot a}(\bar z) \to - \tilde S^{\dot a}(\bar z)~, 
\quad \widetilde{|I\R} \to \widetilde{|I\R}~, 
\quad \widetilde{|a\R} \to - \widetilde{|a\R}~.
\label{Ft-action}
\eea 
Therefore $\tilde S^{\dot a}(\bar z)$ is an NS fermion in the twisted
sector. The NS ground state $\widetilde{|0\R}$ is taken to be odd under
$(-1)^{\Ft}$. Otherwise the set of invariant twisted sector
states reduces to null after imposing the left-right level-matching 
condition. The allowed states in the orbifolded theory are given by,
\bea
\hbox{untwisted} &:& \lt\{ \hbox{any number of $S^a_{-n}$'s} 
\pmatrix{|I\R \cr |\dot a\R} \rt\} \otimes 
\lt\{ \begin{array}{c}
\hbox{even number of $\tilde S^{\dot a}_{-n}$'s} 
\widetilde{|I\R} \\ \\
\hbox{odd number of $\tilde S^{\dot a}_{-n}$'s}\widetilde{ |a\R}
\end{array} \rt\}~, \cr && \cr
\hbox{twisted} &:& \lt\{ \hbox{any number of $S^a_{-n}$'s} 
\pmatrix{|I\R \cr |\dot a\R} \rt\} \otimes \lt\{ \hbox{odd number of 
$\tilde S^{\dot a}_{-r}$'s} \widetilde{|0\R} \rt\}~.
\label{orb-spect}
\eea
It can be shown that,
\bea
\lt\{ \begin{array}{c}
\hbox{even number of $\tilde S^{\dot a}_{-n}$'s} 
\widetilde{|I\R} \\ \\
\hbox{odd number of $\tilde S^{\dot a}_{-n}$'s}\widetilde{ |\dot a\R} 
\end{array} \rt\} \to  \Pi^{(e)}~, \quad
\lt\{ \hbox{odd number of 
$\tilde S^{\dot a}_{-r}$'s} \widetilde{|0\R} \rt\} \to  
\Pi^{(\bar \delta)}~,
\label{orb-spect-two}
\eea
implying that the spectrum in the orbifolded theory is same as that of
the type IIB theory as shown in eq.(\ref{spectIIB}).

\vskip 10pt
\centerline{\it Open String Spectrum}
\vskip 10pt
\noindent
Let us now consider a pair of BPS D-brane and anti-D-brane in type
IIA. There are four sectors of open strings on the
$\hbox{D}\bar{\hbox{D}}$ system with different CP factors
$\Lambda^q$ ($q=1,\cdots , 4$). The various CP factors and the
corresponding open string boundary conditions are give by,
\bea
\Lambda^1= \pmatrix{1 & 0 \cr 0 & 0}: &\quad & 
S^a(\tau, \sigma) =\mbs_{a\dot a} \tilde S^{\dot a}(\tau,\sigma)~, \quad
~~\hbox{at } \sigma =0, \pi~, \cr
\Lambda^2= \pmatrix{0 & 0\cr 0&1}:  & \quad & 
S^a(\tau, \sigma) = - \mbs_{a\dot a} \tilde S^{\dot a}(\tau,\sigma)~, \quad
\hbox{at } \sigma =0, \pi~, \cr
\Lambda^3= \pmatrix{0&1\cr 0&0}: & \quad &
S^a(\tau, 0) =\mbs_{a\dot a} \tilde S^{\dot a}(\tau,0)~, \quad
~~S^a(\tau, \pi) = -\mbs_{a\dot a} \tilde S^{\dot a}(\tau,\pi)~, \cr 
\Lambda^4= \pmatrix{0&0\cr 1 &0}: & \quad & 
S^a(\tau, 0) = - \mbs_{a\dot a} \tilde S^{\dot a}(\tau,0)~, \quad
S^a(\tau, \pi) = \mbs_{a\dot a} \tilde S^{\dot a}(\tau,\pi)~.
\label{opensectors}
\eea
Recall that the matrices $\mbv$, $\mbs$ and $\mbc$ correspond to a BPS
D-brane in the type IIA theory. The above boundary conditions show
that the open strings with CP factors $\Lambda^1$ and
$\Lambda^2$ have integer modding (R sector) and that with CP factors
$\Lambda^3$ and $\Lambda^4$ have half integer modding (NS sector). 
Under the action of $(-1)^{\Ft}$ the above boundary conditions change
following (\ref{Ft-action}). It is easy to see that this action can
alternatively be viewed as an action performed only on the CP
factors. 
\bea
(-1)^{\Ft}: \lt\{ \begin{array}{ll} \Lambda^1 & \leftrightarrow
\Lambda^2 ~, \\ \Lambda^3 &\leftrightarrow \Lambda^4 ~. \end{array} \rt. 
\label{Ft-action-CP}
\eea
Therefore only the R sector states with CP factor 
$\displaystyle{\one_2 \over \sqrt{2}}$
and NS sector states with CP factor $\displaystyle{{\sigma^1\over
\sqrt{2}}}$ survive the projection. Here $\one_2$ and $\sigma^1$ are the
two dimensional identity matrix and the first Pauli matrix
respectively. Therefore the fermionic part of all possible open string 
states living on the non-BPS D-brane of type IIB are given by,
\bea
|\{N_{an}\}, I\R \otimes {\one_2 \over \sqrt{2}} &=& 
\lt( \prod_{a,n} (S^a_{-n})^{N_{an}} \rt) 
|I\R \otimes {\one_2 \over \sqrt{2}}~, \cr
|\{N_{an}\}, \dot a \R \otimes {\one_2 \over \sqrt{2}} &= &
\lt( \prod_{a,n} (S^a_{-n})^{N_{an}} \rt) 
|\dot a \R \otimes {\one_2 \over \sqrt{2}}~, \cr
|\{N_{ar}\}\R \otimes {\sigma^1 \over \sqrt{2}} &= &
\lt( \prod_{a,r} (S^a_{-r})^{N_{ar}} \rt) 
|0\R \otimes {\sigma^1 \over \sqrt{2}}~, 
\label{DDbarspect}
\eea
where $n=1,2, \cdots$, $r=1/2, 3/2, \cdots$, $N_{an}, N_{ar}=0,1$ and 
the product of
oscillators has an implicit fixed ordering. Every state of the above
kind has to
be multiplied with the bosonic part $|\{N_{In}\}, \vec p \R$ to give
the complete state. Here $N_{In}=0,1,2,\cdots$ is the eigenvalue of
the number operator $\al^I_{-n}\al^I_n$ and $\vec p$ is the open string
momentum along the D-brane world-volume. Notice that in the R sector
we have equal number of bosons and fermions at every mass level. This
spectrum is same as that of open strings living on a BPS D-brane of
type IIA. Presence of the NS sector states makes the spectrum
non-supersymmetric.
In particular at the massless level one 
has $|I\R \otimes \one_2/\sqrt{2}$ and $|\dot a\R \otimes
\one_2/\sqrt{2}$ from the R sector. Additional fermionic states 
$|a\R \otimes \sigma^1/\sqrt{2} = S^a_{-1/2} |0\R \otimes
\sigma^1/\sqrt{2}$ come, at this level, from the NS sector showing
that the number of massless fermions is just double compared to
that on a BPS D-brane. This is precisely the case for a non-BPS 
D-brane\cite{bergman98, sen9809, sen99}. Also it is the NS sector
which gives the tachyon $|0\R \otimes \sigma^1/\sqrt{2}$. 

The whole configuration of a non-BPS D-brane in type IIB can be
further projected out by $(-1)^{\Ft}$ of the present theory. In the
bulk it takes us to type IIA theory. On the boundary, this projection
is expected to remove the NS sector states giving the supersymmetric open
string spectrum on a BPS D-brane in type IIA. This can possibly be
shown following the arguments similar to those given in \cite{senrev},
but we have not explicitly gone through that analysis. 

\subsection{Direct Approach}
\label{ss:direct}

Although in the above method the open string theory is simply given
by an orbifold BCFT whose parent BCFT is solved and therefore
potentially does not have any obvious problem, it is
somewhat indirect. Instead of coming from type IIA theory through the 
process of orbifolding, here we shall study the BCFT describing a
non-BPS D-brane directly in type IIB. In this approach the relevant
BCFT should be given by the usual type IIB Green-Schwarz action on the 
world-sheet with a certain open string boundary condition. In the case
of BPS D-branes these boundary conditions can be easily obtained by 
setting the boundary term in the variation of the classical action to zero. 
This can not be done so easily for the non-BPS D-branes\footnote{The
current algebra gluing condition (\ref{Jgluing}) immediately gives
an open string boundary condition in terms of the local $SO(8)$
currents. One can then try to study 
the boundary theory in the current algebraic language. This
formalism requires the additional information of what
kind of highest weight representations are realized in the
theory. Also as we have seen in the boundary state analysis in 
\cite{mukhopadhyay00}, manifest covariance is lost in the current 
algebraic language. Our problem here is to find a covariant
description in terms of the GS fermions.}. 
However, the covariant form
of the boundary state (\ref{nonbpstwo}) allows one to derive certain 
quadratic gluing conditions\footnote{Similar gluing conditions, which
do not exactly match with the results obtained here (see
eq.(\ref{e-gluing-modes})), were found before in 
\cite{nemani}.} which can, in turn, be converted into open string
channel to obtain the required boundary condition. The details of this
analysis is given in appendix \ref{a:bc}. The final result for the
open string boundary condition on the upper half plane (UHP) is given
by,
\bea
S^a(z) S^b(w) &=& \M^{ab}_{cd}~ \St^c(\zb)\St^d(\wb) ~, \quad \quad 
\hbox{at } z=\zb,~w=\wb~.  
\label{Sacondition}
\eea
where,
\bea
\M^{ab}_{cd} &=& {1\over 8} \delta_{ab} \delta_{cd}  + {1\over 16}
\sum_{I,J} \lamb_{\{IJ\}} \g^{IJ}_{ab} \g^{IJ}_{cd} +
{2\over 384} \sum_{\{I,J,K,L\}\in \K} \lamb_{\{IJKL\}}
\g^{IJKL}_{ab} \g^{IJKL}_{cd} ~. \cr &&
\label{M1}
\eea
Equation (\ref{Sacondition}) has to be interpreted to relate normal
ordered operators when the arguments coincide. Let us now explain how 
the above boundary condition is compatible with the condition that one
gets by setting the boundary term in the variation of the action to
zero. In the present case this procedure gives the following condition,
\bea
S^a(z)\delta S^a(z) = \tilde S^b(\zb) \delta \tilde S^b(\zb)~, \quad
\hbox{at } z=\zb.
\label{boundary-variation}
\eea
Contracting the indices on the left hand side of (\ref{Sacondition}) 
and using some of the trace formulas in (\ref{traces}) one finds,
\bea
S^a(z) S^a(w) = \tilde S^b(\zb) \tilde S^b(\wb)~, \quad 
\hbox{at } z=\zb, w=\wb.
\label{scalar-condition}
\eea
The fact that we have bi-local boundary condition enables us to take
variation of the fields at different points on the boundary
independently. We first take the variation of the fields at $w=\wb$ in 
equation (\ref{scalar-condition}). Then after normal ordering the
operators we take the limit: $w\to z$. This gives the condition
(\ref{boundary-variation}) with normal ordering. Notice that it is
difficult to guess the boundary condition (\ref{Sacondition}) from the
requirement of (\ref{boundary-variation}).

Let us now discuss the boundary conditions for the spin fields. For
the vector spin fields this can be readily written down,
\bea
\psi^I(z) = \mbv_{IM} \psit^M(\zb) ~, \quad \hbox{at } z=\zb~.
\label{psicondition}
\eea
Since the conjugate spinor spin field $S^{\dot a}(z)$ appears in the
OPE of $S^a(z)$ with $\psi^I(z)$ (and similarly for the
anti-holomorphic side) any bulk correlation function can be reduced to a
form where only $S^a(z),~\St^a(\zb),~\psi^I(z)$ and $\tilde \psi^I(\zb)$ 
appear. Therefore the boundary conditions 
(\ref{Sacondition}, \ref{psicondition}) are sufficient to compute any such
correlation function. Nevertheless, one can also directly write down the
boundary conditions involving $S^{\dot a}(z)$ and $\St^{\dot a}(\zb)$.
Using the Fiertz identity (\ref{fiertztwo}) one can
decompose $S^{\dot  a}(z)S^{\dot b}(w)$ into antisymmetric tensors of
even ranks. Similarly the identity (\ref{fiertzthree}) enables us to
break $S^{a}(z)S^{\dot b}(w)$ into antisymmetric tensors of odd
ranks. Boundary conditions for these antisymmetric tensors 
will then involve the appropriate twisting given by the matrix $\mbv_{IJ}$. 
The final results can be summarized as,
\bea
\lt. \begin{array}{l}
S^{\dot a}(z) S^{\dot b}(w) = \M^{\dot a\dot b}_{\dot c\dot d}~ 
\St^{\dot c}(\zb)\St^{\dot d}(\wb) ~, \\ \\
S^{a}(z) S^{\dot b}(w) = \M^{a\dot b}_{c\dot d}~ 
\St^{c}(\zb)\St^{\dot d}(\wb) ~, 
\end{array} \rt\} \quad 
\hbox{at } z=\zb,~w=\wb~,  
\label{Sadotcondition}
\eea
where,
\bea
\M^{\dot a\dot b}_{\dot c\dot d} &=& 
{1\over 8} \delta_{\dot a\dot b} \delta_{\dot c\dot d}  + {1\over 16}
\sum_{IJ} \lamb_{\{IJ\}} \gb^{IJ}_{\dot a\dot b} \gb^{IJ}_{\dot c\dot d} +
{2\over 384}  \sum_{\{I,J,K,L\}\in \K} \lamb_{\{IJKL\}}
\gb^{IJKL}_{\dot a\dot b} \gb^{IJKL}_{\dot c\dot d} ~, \cr && \cr
\M^{a\dot b}_{c\dot d} &=& 
{1\over 8} \sum_I \lamb_I \g^{I}_{a\dot b} \g^{I}_{c\dot d} +
{1\over 48}\sum_{IJK} \lamb_{\{IJK\}} \g^{IJK}_{a\dot b} 
\g^{IJK}_{c\dot d} ~. 
\label{M2M3}
\eea
The above boundary conditions look complicated and we have not tried to
directly quantize the corresponding world-sheet theory. Instead we
take an approach where the solution is given in terms of a minimal set
of rules for computing all possible correlation functions in the
theory. In the following we design a special ``doubling trick'' and
use that to compute the open string spectrum and prescribe a set of rules
for computing all possible correlation functions without boundary
insertions leaving the rest of the correlators for future work. 

\subsubsection{Boundary Spectrum and Prescription for Computing 
Correlation Functions without Boundary Insertions} \label{sss:prescription}

In the BCFT describing a BPS D-brane the fundamental world-sheet
fields have linear boundary conditions. In that case one can use
the usual doubling trick \cite{gubser96} to reduce a correlation
function of bulk operators on the UHP to a correlation function on 
the full plane which is simply a computation on the holomorphic side 
of the bulk theory without boundary. As we have seen above, in case 
of a non-BPS D-brane the world-sheet fields that are space-time spinors
have bi-local boundary conditions. In this case we generalize the
above-mentioned trick to achieve the same thing.   

The local properties of the world-sheet fields are same both in the
boundary and the bulk theory. Let us consider a set of three
holomorphic fields 
$\lt\{\CS^a(z),\CS^{\dot a}(z),\Psi^I(z)\rt\}$
with the implied $SO(8)$ transformation properties living on the full 
complex plane which have the same local properties as the holomorphic
triad $\lt\{ S^a(z),S^{\dot a}(z),\psi^I(z) \rt\}$
of the bulk theory. As usual, we define $\Psi^I(z)$ to be,
\bea
\Psi^I(u) = \lt\{ \begin{array}{ll}
\psi^I(z)|_{z=u}~, & \quad \Im u \geq 0~,\\ & \\
\lamb_I \psi^I(\zb)|_{\zb=u}~, & \quad \Im u \leq 0~. \\
\end{array} \rt.
\label{Psi}
\eea
We actually intend to consider bi-local operators constructed out of 
$\CS^a(z)$ and $\CS^{\dot a}(z)$ rather than the individual
ones. These are given by,
\bea
\begin{array}{lll}
\CS^a(u) \cdots \CS^b(v) =& \lt\{ \begin{array}{l} 
S^a(z)\cdots S^b(w)|_{z=u,w=v}~, \\  \\
\M^{ab}_{cd}~ \St^c(\zb)\cdots \St^d(\wb)|_{\zb=u, \wb=v}~, 
\end{array} \rt. &
\begin{array}{l}
\Im u, \Im v \geq 0~, \\ \\ \Im u, \Im v \leq 0~, \end{array} 
\end{array}
\label{CSaCSb}
\eea 
\bea
\begin{array}{lll}
\CS^{\dot a}(u) \cdots \CS^{\dot b}(v) =& \lt\{ \begin{array}{l} 
S^{\dot a}(z)\cdots S^{\dot b}(w)|_{z=u,w=v}~, \\  \\
\M^{\dot a\dot b}_{\dot c\dot d}~ \St^{\dot c}(\zb)\cdots 
\St^{\dot d}(\wb)|_{\zb=u, \wb=v}~, \end{array} \rt. &
\begin{array}{l}
\Im u, \Im v \geq 0~, \\ \\ \Im u, \Im v \leq 0~, \end{array} 
\end{array}
\label{CSadotCSbdot}
\eea 
\bea
\begin{array}{lll}
\CS^a(u) \cdots \CS^{\dot b}(v) =& \lt\{ \begin{array}{l} 
S^a(z)\cdots S^{\dot b}(w)|_{z=u,w=v}~, \\  \\
\M^{a\dot b}_{c\dot d}~ \St^c(\zb)\cdots \St^{\dot d}(\wb)|_{\zb=u,  \wb=v}~, 
\end{array} \rt. &
\begin{array}{l}
\Im u, \Im v \geq 0~, \\ \\ \Im u, \Im v \leq 0~, \end{array} 
\end{array}
\label{CSaCSbdot}
\eea 
The dots imply that the above relations are understood to be true even
when other operators appear in between in a correlation function.
Notice also that both the arguments $u$ and $v$ are either in the
upper or the lower half plane. This implies that we need to consider
only a sub-sector of all possible holomorphic correlation functions 
of $\CS^a(z)$, $\CS^{\dot a}(z)$ and $\Psi^I(z)$ on the full plane. 

Having defined the essential fields on the full plane, we can now
derive the boundary spectrum quite easily\footnote{I thank
A. Sinha and N. V. Suryanarayana for discussion on this point.}. 
Using eq.(\ref{CSaCSb}) and the boundary condition (\ref{Sacondition}) 
one can show that on the cylinder,
\bea
{\cal S}^a(\tau, 2\pi){\cal S}^b(\tau^{\prime}, 2\pi) &=&
{\cal S}^a(\tau, 0){\cal S}^b(\tau^{\prime}, 0)~, \cr
\hbox{i.e. }  {\cal S}^a(\tau, 2\pi) &=& \pm {\cal S}^a(\tau, 0)~,
\label{CSacondition}
\eea
which says that there is an R and an NS sectors of states in the
boundary spectrum in agreement with the result found in
subsec.\ref{ss:orbifold}. 

Assigning the fermion numbers $F$ and $\Ft$, defined only mod 2,
for the left and right-moving fields as shown in table \ref{Fno},
\begin{table}[t]
\centerline{
\begin{tabular}{|c|c|c|c|c|c|c|}
\hline 
  & $S^a(z)$ & $\St^a(\zb)$ & $\psi^I(z)$ & $\psit^I(\zb)$ & 
$S^{\dot a}(z)$ & $\St^{\dot a}(\zb)$ \\ \hline
$F$ & $1$ & $0$ & $0$ & $0$ & $1$ & $0$ \\ \hline
$\Ft$ & $0$ & $1$ & $0$ & $0$ & $0$ & $1$ \\
\hline
\end{tabular}}
\caption{Assignment of fermion numbers to various bulk fields}
\label{Fno}
\end{table} 
we now give our prescription for computing a generic correlation function
of the bulk insertions on UHP.

\begin{enumerate}
\item
In order for the correlation function to be nonzero the net global
fermion charges of all the closed string insertions should be
$F_C=\tilde F_C=0$. This implies that if $m,n,p,q,r,s$ are the 
number of $S^a(z)$'s, $\St^a(\zb)$'s, $S^{\dot a}(z)$'s, 
$\St^{\dot a}(\zb)$'s, $\psi^I(z)$'s and $\psit^I(\zb)$'s
respectively, then the correlator is nonzero only when 
\bea
m+p = \hbox{even}~, \quad n+q = \hbox{even}~, \quad 
r, s~\hbox{arbitrary }.
\label{F-charge}
\eea
\item
A correlation function satisfying the conditions in  (\ref{F-charge}) 
can be computed with the following prescription.

\begin{itemize}
\item
Pair up all the left and right moving fields that are space-time
spinors such that all of them can be written in terms of the bilinear
of $\CS^a(z)$ and $\CS^{\dot a}(w)$ by using eqs.(\ref{CSaCSb},
\ref{CSadotCSbdot}, \ref{CSaCSbdot}). Do-ability of this is guaranteed 
by the condition (\ref{F-charge})\footnote{Generically this can be
done in more than one ways. Consistency would require all of them to 
give the same final result. Although we expect that it should be
possible to establish this using various trace formulas in
(\ref{traces}), we have not found a complete proof.}. 
\item
Using the usual doubling trick write all the left and right moving
fields that are space-time vector in terms of $\Psi^I(z)$'s. 
\item
Performing the above two steps we end up having a holomorphic
correlation function on the full plane. Compute this following the
same rules for computing the holomorphic correlation functions of the
bulk theory without boundary.
\end{itemize} 
\end{enumerate}
The second rule described above is simply a generalization of the
usual doubling trick to the present case. It is difficult to 
compute the correlators which do not satisfy condition (\ref{F-charge}). 
Although we do not have a direct proof of the first rule, it gives the
expected result that a non-BPS D-brane does not source RR closed
string states. Notice that the corresponding statement for a BPS
D-brane is that a correlator without boundary insertions is nonzero 
only when $F_C + \tilde F_C =0$. 

\subsubsection{Closed String One-point Functions}
\label{sss:onepoint}

Here we shall demonstrate through explicit computations how to use the
above prescription to find bulk correlators. In particular we shall
compute several closed string one-point functions on the unit disk and show
that when the vertex operator is inserted at the center of the disk
the result matches with that obtained through boundary state computation.

We shall verify the following relation for certain closed string states
$|\Phi\R$,
\bea
\langle \langle e | \Phi \R = C \lim_{\zeta, \bar \zeta  \to 0} 
\langle \Phi(\zeta ,\bar \zeta ) \R_D~,
\label{one-point}
\eea
where $\Phi(\ze, \zeb)$ is the vertex operator for the state $|\Phi\R$
with $(\ze , \zeb)$ being the complex coordinate system on the unit
disk. $\langle \cdots \R_D$ denotes a disk correlation function in the BCFT
of fermions that we are presently considering. $C$ is an
overall constant that does not depend on which closed string state we choose.
The states for which we verify relation (\ref{one-point}) are,
\bea
|\Phi^{IJ}\R &=& |I\R \otimes \widetilde{|J\R} ~, \cr && \cr
|\Phi^{IJ}_{m,n}\R &=& - \g^I_{a\dot a} \g^J_{b\dot b} S^a_{-m}
\St^b_{-n} |\dot a\R \otimes \widetilde{|\dot b\R} ~, \cr && \cr
|\Phi^{(IJK),(MNP)}_{m,n}\R &=& - \g^{IJK}_{a\dot a}
\g^{MNP}_{b\dot b} S^a_{-m} \St^b_{-n}
|\dot a\R \otimes \widetilde{|\dot b\R} ~,\cr && \cr
|\Phi^{IJ}_{m,n,p,q}\R &=& J_{-m,-n}\Jt_{-p,-q}|I\R \otimes
\widetilde{|J\R} ~, \cr && \cr
|\Phi^{(MN),(M^{\prime}N^{\prime}),IJ}_{m,n,p,q}\R &=& J^{MN}_{-m,-n}
\Jt^{M^{\prime}N^{\prime}}_{-p,-q}|I\R \otimes
\widetilde{|J\R}~, \cr && \cr
|\Phi^{(MNPQ),(M^{\prime}N^{\prime}P^{\prime}Q^{\prime}),IJ}_{m,n,p,q}\R
&=&
J^{MNPQ}_{-m,-n}\Jt^{M^{\prime}N^{\prime}P^{\prime}Q^{\prime}}_{-p,-q}
|I\R \otimes \widetilde{|J\R}~.
\label{states}
\eea
These classes of states have been chosen to explicitly verify the numerical
factors $1/8$, $1/16$, $2/384$ and $1/48$ appearing in
eqs.(\ref{XM}, \ref{YM}).
The above states are antisymmetric under interchanges of the vector
indices that are kept inside $(\cdots)$. To demonstrate the
computation, here we shall verify eq.(\ref{one-point}) only for the first two
classes of states in the above list leaving the analysis for the rest
in appendix \ref{a:onepoint}. The vertex operators for the states
$|\Phi^{IJ}\R $ and $|\Phi^{IJ}_{m,n}\R $ are,
\bea
\Phi^{IJ}(\ze,\zeb) = \psi^I(\ze) \psit^J(\zeb)~, \quad
\Phi^{IJ}_{m,n}(\ze,\zeb) = - \g^I_{a\dot a} \g^J_{b\dot b}
\Phi^{ab \dot a  \dot b}_{m,n}(\ze,\zeb) ~,
\label{tensor-vertices-demo}
\eea
respectively, where,
\bea
\Phi^{ab \dot a \dot b}_{m,n}(\ze,\zeb) &=& - \oint_{\ze}
{d\ze_1 \over 2\pi i} \oint_{\zeb} {d\zeb_2 \over 2\pi i}
(\ze_1-\ze)^{-m-1/2} (\zeb_2-\zeb)^{-n-1/2}
S^a(\ze_1) \St^b(\zeb_2) S^{\dot a}(\ze) \St^{\dot b}(\zeb) ~.\cr &&
\label{spinor-vertex1}
\eea
Defining the bra state: $\langle\Phi|=\langle\phi_R|\otimes \langle\phi_L|$
corresponding to a ket $|\Phi\R = |\phi_L\R \otimes |\phi_R\R$, where
$\langle\phi_{R/L}|$ has the conjugated oscillators placed just in the
reverse order of that in $|\phi_{R/L}\R$, we get
the following inner products,
\bea
\langle \Phi^{IJ}|\Phi^{I^{\prime} J^{\prime}} \R = \dt_{I,I^{\prime}}
~\dt_{J,J^{\prime}}~, \quad
\langle \Phi^{IJ}_{m,n} | \Phi^{I^{\prime}
J^{\prime}}_{m^{\prime},n^{\prime}} \R = 64 \dt_{m,m^{\prime}}
~\dt_{n,n^{\prime}}~\dt_{I,I^{\prime}}  ~\dt_{J,J^{\prime}} ~.
\label{inner-product-demo}
\eea
States belonging to different classes in eqs.(\ref{states}) can be
shown to be mutually orthogonal. Using these inner products and
eq.(\ref{nonbpstwo}) we get the following results,
\bea
\langle \langle e |\Phi^{IJ} \R = \lamb_I \delta_{IJ} ~, \quad
\langle \langle e |\Phi^{IJ}_{m,n} \R = 8 \dt_{m,n}~\lamb_I \delta_{IJ} ~.
\label{bstate-results-demo}
\eea
We shall now compute the right hand side of eq.(\ref{one-point}) for
the vertex operators in eqs.(\ref{tensor-vertices-demo}) in the BCFT using
the prescriptions in \ref{sss:prescription}. Using the following
conformal transformation which relates the unit disk $(\ze, \zeb)$ to UHP
$(z, \zb)$,
\bea
z(\ze) = i {1+\ze \over 1- \ze}~,
\label{conf-tfn}
\eea
one can relate correlation functions of bulk primaries on unit disk
and on UHP in the following way,
\bea
\langle \prod_i \Phi_i(\zeta_i, \bar \zeta_i) \R_D &=&
\prod_i \lt((z_i^{\prime})^{h_i} (\bar z_i^{\prime})^{\bar h_i}\rt)
\langle \prod_i \Phi_i(z_i, \bar z_i)\R_{UHP} ~,
\label{disk-UHP}
\eea
where $\Phi_i$ is a bulk primary with conformal weight $(h_i,\bar
h_i)$, $z_i=z(\zeta_i)$ and $z^{\prime}_i=z^{\prime}(\zeta_i)$ with
the prime denoting the first derivative. Using this and the standard doubling
trick for the vertex $\Phi^{IJ}(\ze, \zeb)$ one gets,
\bea
\langle \Phi^{IJ}(0,0) \R_D &=& 2 \langle \psi^I(i) \psit^J(-i) \R_{UHP}~,\cr
&=& 2 \lamb_J \langle \Psi^I(i) \Psi^J(-i) \R~,
\eea
where $\langle \cdots \R$ refers to a holomorphic correlation function on
the full plane. Finally using the OPE:
$\Psi^I(z) \Psi^J(w) \sim {\dt_{IJ} \over (z-w) }$ one gets,
\bea
\langle\Phi^{IJ}(0,0)\R_D = \lamb_I \delta_{IJ}~.
\label{d1}
\eea
Comparing this with the first equation of (\ref{bstate-results-demo}) one
fixes the constant:
\bea
C= 1~.
\label{const}
\eea
We then turn to the computation of $\langle \Phi^{IJ}_{m,n}(0,0)
\R_D$. Using (\ref{disk-UHP}) we may write,
\bea
\langle \Phi^{IJ}_{m,n}(0,0) \R_D = - \oint_i {dz_1 \over 2\pi i}
\oint_{-i} {d \bar z_2 \over 2\pi i} {\cal J}_{m,n}(z_1, \bar z_2)
\g^I_{a\dot a} \g^J_{b\dot b}
\langle S^a(z_1) \St^b(\bar z_2) S^{\dot a}(i) \St^{\dot
  b}(-i) \R_{UHP}~,
\label{one-pointIJmn1}
\eea
where,
\bea
{\cal J}_{m,n}(z_1, \bar z_2) =
4 (z_1-i)^{-m-1/2} (z_1+i)^{m-1/2} (\bar z_2+i)^{-n-1/2}
(\bar z_2-i)^{n-1/2} ~.
\label{Jacobian-mn}
\eea
Using the generalized doubling trick described in the previous
subsection we now reduce the right hand side of eq.(\ref{one-pointIJmn1})
to a correlation function on the full plane,
\bea
\langle \Phi^{IJ}_{m,n}(0,0) \R_D = \lamb_J \oint_i {dz \over 2\pi i}
\oint_{-i} {dw \over 2\pi i} {\cal J}_{m,n}(z, w)
\g^I_{a\dot a} \g^J_{b\dot b}
\langle \CS^a(z) \CS^b(w) \CS^{\dot a}(i) \CS^{\dot  b}(-i) \R~.
\label{disk0}
\eea
Notice that there is an overall sign difference between the
eqs. (\ref{one-pointIJmn1}) and (\ref{disk0}). This is simply because
the second integral in eq.(\ref{one-pointIJmn1}) has been converted to
a holomorphic integral in eq.(\ref{disk0}) which reverses the sense of
the contour.
The correlation function in the above integrand and all others that are
needed for the computations done in  appendix \ref{a:onepoint}
have been evaluated in appendix \ref{a:correlators}. Using
eq.(\ref{corr0}) and the result for the relevant integral given in
eq.(\ref{int1}) one gets,
\bea
\langle \Phi^{IJ}_{m,n}(0,0) \R_D = 8  \delta_{mn} \lamb_I \delta_{IJ}~.
\eea
With $C=1$ and the second equation in (\ref{bstate-results-demo}) the
above result verifies (\ref{one-point}) for the states $|\Phi^{IJ}_{m,n}\R$.

\sectiono{Discussion}
\label{s:discussion}

Here we discuss a few points that are relevant to the present work
including some future directions.

\begin{itemize}
\item
{\bf Method of Ref.\cite{nemani}}

It was suggested in \cite{nemani} that the covariant expression for a
non-BPS boundary state can be obtained by going through the following
two steps:
\begin{enumerate}
\item
Write the NSNS part of a BPS boundary state (whose covariant
expression is known) in a form where spinor and conjugate spinor
matrices $\ms$ and $\mc$ respectively do not appear at all, rather
only the vector representation matrix $\mv$ appears.
\item
Replace $\mv$, which corresponds to a ``BPS automorphism'' by $\mbv$,
the one corresponding to a ``non-BPS automorphism'', in the
final expression obtained in the first step.
\end{enumerate}
This method is expected to work because of the following reason.
Notice that being $SO(8)$ tensors the current modes are glued with the
vector representation matrix $\mv$ or $\mbv$, depending on whether
the relevant D-brane is BPS or non-BPS respectively
(see eqs.(\ref{Jgluing}, \ref{outerauto})). Therefore it
should be possible to write the coefficient of any term in the basis
expansion of the Ishibashi state purely in terms of the vector
representation matrix. This, along with the fact that both the NSNS
part of a BPS boundary state and a non-BPS boundary state
belong to the same subspace ($\Pi^{(e)}\otimes \Pi^{(e)}$) of the full
closed string Hilbert space
\cite{mukhopadhyay00}, guarantees the do-ability of the first step.
Once the first step is achieved the second step is analogous to
computing a function for a different argument.

This method works well
provided a particular technical subtlety is taken into account. After
completing the first step if one naively goes through the replacement in
the second step then one finds that the coefficients of a class of
basis states in the expansion of the boundary state become zero which
were previously nonzero. Similar situation arises in the current
algebraic derivation given in appendix \ref{a:CAderivation} as well.
Although the present method by itself does not say anything special
about this situation, we have argued in appendix \ref{a:CAderivation}
(see discussion below eq.(\ref{TonX})) that this is unexpected. To
arrive at the expression (\ref{nonbpstwo}) one has to take care of
this subtlety in a way similar to that prescribed in appendix
\ref{a:CAderivation}. Recall that it is the expression
(\ref{nonbpstwo}) which passes through the open-closed duality check
in appendix \ref{a:duality}.

\item
{\bf Comments of the computation of open-point functions}

As mentioned below eq.(\ref{gammas}), the overall sign of the state
$|\mbv, \odd \R$ in eqs.(\ref{nonbpstwo}) is convention dependent. Had
we followed the
convention of \cite{mukhopadhyay00} (see footnote \ref{convention})
this sign would have been $(+)$. Notice that the sign of the state 
$|\mbv, \ev \R$ does not depend on the convention. One may wonder
how to see this ambiguity in the computation of the one-point
functions on the BCFT side. In the BCFT method we have reduced the 
computation of the one-point function of a closed string state to 
computing certain complex integrals where a holomorphic correlation
function on the full plane appears in the integrand. Because of the 
presence of the spin fields the results of these correlation functions 
involve various branch cuts on the complex plane. In order to compute 
the integrals the correlation functions have to be computed for
certain ranges of values of the arguments. Typically this procedure is 
ambiguous as one has to make choice of the branches. It turns out that 
this feature affects our final result (by introducing a sign ambiguity)
only for the second and third classes of the states in the list 
(\ref{states}), which are precisely the ones relevant to the state 
$|\mbv, \odd\R$. This can be seen in the following way: 
The contours for the required integrals have been shown in
fig.\ref{fig:contours}(A). This can be computed by taking the limit:
$u\to i,~ v\to -i$ on the contours in fig.\ref{fig:contours}(B). The 
integrals relevant to fig.\ref{fig:contours}(B) requires one to
compute the relevant correlators, as given by
eqs.(\ref{corr0}, \ref{corr1}), in a region where $|u|>|w|$. This,
in turn, requires us to pass $\CS^b(w)$ through $\CS^{\dot
a}(u)$. This procedure contributes a factor of $(\pm i)$, as evident
from the square root brunch cut in $(w-u)$ present in the results 
(\ref{Ftld11}, \ref{Ftld33}). It can be checked that the other part of
the integrand, namely ${\cal J}_{mn}(z,w,u,v)$ (such that ${\cal
  J}_{mn}(z,w,i,-i) = {\cal J}_{mn}(z,w)$ which is given in
eq.(\ref{Jacobian-mn})) does not have a brunch cut in $(w-u)$. After taking
the required limit all the brunch cuts go away by absorbing the factor
of $i$, but leaving a sign ambiguity. Notice that this
sign ambiguity does not arise for the other correlators in
(\ref{corr2}, \ref{corr3}, \ref{corr4}) which are relevant for 
computing one-point functions of some of the states in $|\mbv, \ev
\R$. It can be checked from the results given in
eqs.(\ref{F11}, \ref{F331}, \ref{F332}, \ref{F551}, 
\ref{F552}, \ref{F1F2F3}) that no extra phase arises while passing
$z_2$ and $w_2$ simultaneously through $u$. 
\begin{figure}
\begin{center}
\leavevmode
\hbox{%
\epsfxsize=4.0in
\epsfysize=3.0in
\epsffile{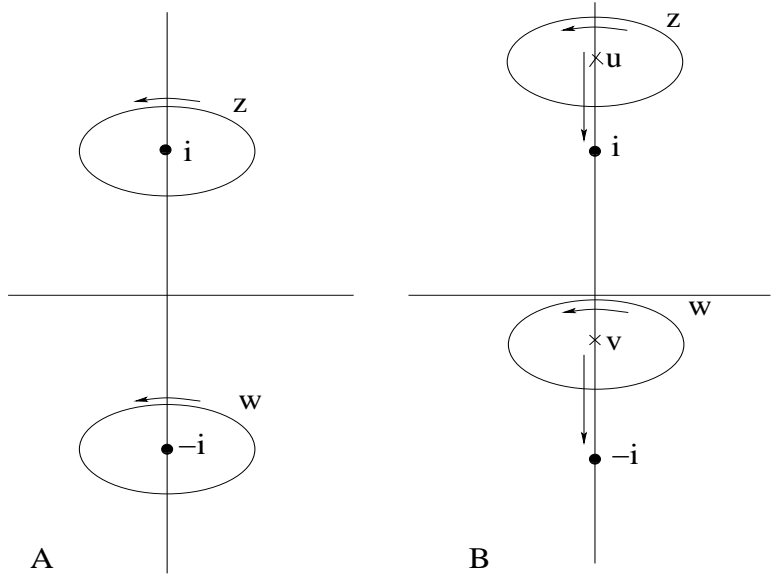}}
\caption{(A) Contours for computing the integrals relevant to the one-point
functions of the second and third classes of states in the list
(\ref{states}). (B) A possible limiting procedure to compute the
integrals.}
\label{fig:contours}
\end{center}
\end{figure}
\item
{\bf More on BCFT}

Both the approaches in sec.\ref{s:BCFT} have been given partial
description. We have not discussed computation of the correlation
functions in the ``orbifold approach''. Although this is an orbifold
BCFT whose parent theory is solved and one does not see any obvious
problem in computing correlators, it may be useful to understand the
details of the computations. We have prescribed rules for computing
the correlators without boundary insertions in the ``direct
approach''. One might get clues on how to set up rules for computing
correlators with boundary insertions by going through those
computations in the ``orbifold approach''.  
\item
{\bf Trial in pp-wave}

We have given some naive attempts to find the analogous non-BPS
D-barnes in the type IIB pp-wave background. There is a lot of
algebraic similarity between string quantization in this and in the
flat background \cite{metsaev01}. In fact the fermionic parts of the
boundary states 
of class I (according to the classification of \cite{gaberdiel02}) 
D-branes in pp-wave\cite{billo02, bergman02} look exactly same as that
of the BPS D-branes in flat space (see, for example, eqs.(2.13) and
(2.14) in \cite{bergman02}). Given 
this one may wonder if a class of non-BPS boundary states in pp-wave 
can be obtained by taking the fermionic parts to be exactly same as 
those (eqs.(\ref{nonbpstwo})) in the flat background. This is also
precisely what one would do if one generalizes the method of
\cite{nemani}, as described above, to pp-wave background. It turns out
that one can use the same bosonization and refermionization technique 
of sec.\ref{s:BS} to compute the necessary overlap of the boundary
sate to evaluate the cylinder diagram. This is found not to have an 
open string interpretation unless the BPS condition is
satisfied, namely $|r-s|=2$, where $r$ and $s$ are the number of
Neumann directions in first and last four coordinate directions 
$(x^1, \cdots, x^4)$ and $(x^5, \cdots , x^8)$ respectively. Clearly
this can not be satisfied for the case that we are interested in,
namely $r+s=$ odd.

One may try another generalization from the flat space to pp-wave in
the following way. We have shown in sec.\ref{s:BS} that any non-BPS 
boundary state (fermionic part) in flat space can be given by the NSNS
part of a BPS D-instanton boundary state written in terms of the
oscillators of $S^a(z)$ and $\bar S^a(\zb)$ (see eqs.(\ref{Shat},
\ref{Shat-Stld}, \ref{Sbar-Shat})). The information about the 
dimensionality and alignment of the brane are completely absorbed into
the definition of $\bar S^a(\zb)$. 
Therefore if the NSNS part of a BPS D-instanton boundary sate
(eqs.(2.15)-(2.17) of \cite{gaberdiel02})
written in terms of the oscillators of $S^a(z)$ and $\bar S^a(\zb)$
admits an open string interpretation in pp-wave then we are done. The
answer turns out to be ``no''.

All these attempts are at a very naive level where one tries to
exploit some algebraic features in the string quantization. Notice
that the algebraic structure of the string zero modes in pp-wave is
quite different from that in flat space. In fact the failure in both 
the above computations is actually caused by the fermion zero modes.
It was pointed out in \cite{bergman02} that given a form of a boundary
state, these zero modes put strong restriction on the dimensionality
and alignment of the D-brane. Indeed the boundary states have been
classified in \cite{gaberdiel02} on the basis of dimensionality and 
alignment of D-branes. The reason why the above methods do not work
may be that we are trying to enforce a particular form of the boundary
state for a D-brane of different dimensionality. It might very well 
happen that the non-BPS D-branes that we are looking for do exist with
some completely different forms of the boundary states. If possible,
it will be very useful to have a current algebraic formulation where
this question may be asked more clearly.

\end{itemize}
\medskip
\centerline{\bf Acknowledgement}
\noindent
I am thankful to A. Sinha and N. V. Suryanarayana for many useful
discussions. I would like to thank N. V. Suryanarayana also for
communicating his work in \cite{nemani}. This attempt has encouraged
the present study to a great extent. I also thank S. R. Das,
M. Lippert, J. Majumder, J. Michelson, A. Shapere, X. Wu and C. Zhou for
discussions. A few very helpful conversations with A. Sen are also
thankfully acknowledged. I apologize to every author whose work was
not cited in the previous version of this paper. This work was
supported by the DOE grant DE-FG01-00ER45832.

\appendix

\sectiono{Current Algebraic Construction}
\label{a:CAconstruction}

While the perturbative excitation spectrum (only the world-sheet
fermionic part) of type IIA theory is given by eq.(\ref{spectIIA}),
the same for type IIB theory can be shown to be, 
\bea
\hbox{type IIB}: \quad (\Pi^{(e)} \oplus \Pi^{(\bar \delta)}) \otimes
(\Pi^{(e)} \oplus \Pi^{(\bar \delta)}) ~. 
\label{spectIIB}
\eea
The above notation has been explained below eq.(\ref{spectIIA}). 
It was argued in \cite{mukhopadhyay00} that the fermionic part 
$|e\R\R$ of the non-BPS boundary state in eq.(\ref{nonbps}) is the 
Ishibashi state corresponding to the highest weight $e$ and it
satisfies the following gluing condition,
\bea
\lt (J^{IJ}_n + \tau . \tilde J^{IJ}_{-n} \rt) |e\R \R = 0~, \quad
\forall n \in \ZZZ~, 
\label{Jgluing}
\eea
where $J^{IJ}_n$ and $\tilde J^{IJ}_n$ are the modes of the local left
and right moving $SO(8)$ currents respectively constructed out of the
world-sheet fermions. 
\bea
J^{IJ}_n = {i\over 4} \lt(\g^I \gb^J \rt)_{ab} \sum_{m\in \ZZZ} 
:S^a_m S^b_{n-m}:~, \quad \quad \forall n \in \ZZZ~,
\label{Jmodes}
\eea
(similarly for the right moving current modes) where $::$ denotes the
usual oscillator normal ordering. $\tau$ is the outer automorphism given 
by\footnote{These outer automorphisms which correspond to reflections
along odd number of directions take the vector representation to
itself but switches between the spinor and conjugate spinor
representations. This is why only the Ishibashi state $|e\R \R$ is
realized in a non-BPS boundary state. Notice that in the BPS case the
automorphisms
involved are inner as they correspond to reflections along even number
of directions. In this case all the representations go to
themselves. As a result both the Ishibashi states $|e\R \R$ and
$|\bar \delta \R \R$ are realized forming the NSNS and the RR parts
of the boundary state respectively \cite{mukhopadhyay00}.},
\bea
\tau . \tilde J^{IJ}_n = \mbv_{IK} \mbv_{JL} \tilde J^{KL}_n~.
\label{outerauto}
\eea
The Ishibashi state $|e\R \R$ is given by,
\bea
|e \R \R = \sum_{N} |N, e \R \otimes {\cal T} \Theta \widetilde{
|N, e\R} ,
\label{e}
\eea
where $|N,e\R$ is a complete set of orthonormal basis vectors in 
$\Pi^{(e)}$:
\bea
\lt\{|N,e \R \rt\} \equiv 
\lt\{ \lt[\prod_{a,n>0} \lt(S^a_{-n} \rt)^{N_{an}} |I \R 
\rt]_{\sum N_{a n} =even}~, \quad
\lt[\prod_{a,n>0} \lt( S^a_{-n}  \rt)^{N_{an}} |\dot a\R 
\rt]_{\sum N_{an} =odd} \rt\} ~, 
\label{basis}
\eea
where $N_{an}=0,1$ and the products are ordered products with some
chosen ordering.
$\T$ and $\Theta$ are two Hilbert space operators which act only on
the right moving part of a state. It was argued in
\cite{mukhopadhyay00} that the action of $\Theta $ on these basis
states is actually trivial, i.e.
\bea
\Theta \widetilde{|N,e\R} = \widetilde{|N,e\R}~.
\label{actTheta}
\eea
All the states in (\ref{basis}) can be obtained by applying current 
creation operators on the highest weight state 
$|e\R = (|I=1\R - i |I=2\R)/\sqrt{2}$. The action of $\T$ on all such
states can be obtained from the following,
\bea
\T \tilde J^{IJ}_n \T^{-1} &=& \tau . \tilde J^{IJ}_n = 
\mbv_{IK} \mbv_{JL} \tilde J^{KL}_n ~,\cr
\T \widetilde{|I\R} &=& \mbv_{IJ} \widetilde{|J\R} ~,
\label{Taction}
\eea
Although we have a nice action of $\T$ on the current oscillators,
being an outer automorphism, $\T$ does not have a well-defined action on
the $\tilde S^a_n$ oscillators. This is the main obstruction against
finding a simple covariant expression for $|e\R \R$. 

\sectiono{Derivation of the Covariant Form}
\label{a:CAderivation}

Despite the problem mentioned below eq.(\ref{Taction}), the covariant 
expression for non-BPS boundary state given in
eqs.(\ref{nonbpstwo}, \ref{XM}, \ref{YM}) can be derived in the current
algebraic framework by manipulating the basis states (\ref{basis}) in
a certain manner.
Taking the expression of $|e\R \R$ given in (\ref{e}) and using the
basis states (\ref{basis}) one gets,
\bea
|e\R \R &=& |\ev \R + |\odd \R ~, \cr
|\ev \R &=&  {1\over 2}\T \sum_{\{N_{an}\}, I} \lt(1+
(-1)^{\sum_{a,n}N_{an}} \rt) |\{N_{an}\}, I\R ~, \cr
|\odd \R &=& {1\over 2}\T \sum_{\{N_{an}\}, \dot a} \lt(1-
(-1)^{\sum_{a,n}N_{an}} \rt) |\{N_{an}\}, \dot a\R ~,
\label{CAe}
\eea
where,
\bea
|\{N_{an}\}, I \R &=& \prod_{n>0,a} \lt(S^a_{-n}\rt)^{N_{an}} |I\R
\otimes \prod_{n>0,a} \lt(\tilde S^a_{-n}\rt)^{N_{an}} \widetilde{|I\R} ~, \cr
&=& \prod_{n>0,a} \lt(i S^a_{-n} \tilde S^a_{-n} \rt)^{N_{an}} 
|I\R \otimes \widetilde{|I\R} ~. 
\label{stateI}
\eea
\bea
|\{N_{an}\}, \dot a \R &=& \prod_{n>0,a} \lt(S^a_{-n}\rt)^{N_{an}}
|\dot a\R \otimes \prod_{n>0, a} \lt(\tilde S^a_{-n}\rt)^{N_{an}}
\widetilde{|\dot a\R} ~, \cr
&=& i (-1)^{1+\sum_{a,n}N_{an}}  \prod_{n>0,a} \lt(i S^a_{-n} 
\tilde S^a_{-n} \rt)^{N_{an}} 
|\dot a\R \otimes \widetilde{|\dot a \R}~. 
\label{stateadot}
\eea
Notice that while going from the first step to the second in
eq. (\ref{stateadot}) we need to keep in mind that each $\tilde S^a_{-n}$
oscillator picks up a $(-)$ sign when it passes through the state
$|\dot a\R$. Plugging in the result (\ref{stateI}) into the second 
equation of (\ref{CAe}) one can write,
\bea
|\ev\R &=& \T \cos \lt(\sum_{n>0,a} S^a_{-n} \tilde S^a_{-n} \rt) |I\R
\otimes \widetilde{|I\R } ~, \cr
&=& \T \cosh\lt( \sqrt{X}\rt) |I\R \otimes \widetilde{|I\R} ~,
\label{even}
\eea
where,
\bea
X &=& \sum_{m,n>0} S^a_{-m}S^b_{-n} \tilde S^a_{-m} \tilde S^b_{-n}~, \cr
&=& \sum_{m,n>0} \lt[{1\over 8} J_{-m,-n} \tilde J_{-m,-n} + 
{1 \over 16} J^{IJ}_{-m,-n} \tilde J^{IJ}_{-m,-n} + 
{1 \over 384} J^{IJKL}_{-m,-n} \tilde J^{IJKL}_{-m,-n} \rt]~.
\label{X}
\eea
where the second line (with the definition (\ref{Jmn}) for both the
left and right moving variables) can be derived by using the Fiertz
identity (\ref{fiertzone}) and some of the trace formulas in 
(\ref{traces}). As we have gotten rid of
the spinor indices we may now hope to evaluate the action of $\T$,
\bea
\T X \T^{-1}= X_{\mbv} &=& \sum_{m,n>0}
\lt[{1\over 8} J_{-m,-n} \tilde J_{-m,-n} + 
{1\over 16} \sum_{IJ} \lamb_{\{IJ\}} J^{IJ}_{-m,-n} \tilde J^{IJ}_{-m,-n} + 
\rt. \cr
&& \lt. {1\over 384} \sum_{IJKL} \lamb_{\{IJKL\}} J^{IJKL}_{-m,-n} 
\tilde J^{IJKL}_{-m,-n} \rt]~.
\label{TonX}
\eea
The last term on the right hand side with free summations over all the 
four indices is zero because of eq.(\ref{lamb-duality})
and the self-duality of $\tilde J^{IJKL}_{m,n}$ (eq.(\ref{self-dualJ})).
There are various ways to see that this is not correct. 
One can study the non-BPS boundary state in NSR formalism and
explicitly check that the states that are created by this term are
present. One can also argue in the current algebraic framework.
If this term were zero then it would have implied that $\T$
annihilates some of the states in $\Pi^{(e)}$. On the other hand since
$\lt(\mbv \rt)^2=\one_8$, $\T^2$ should be expected to be identity. 
This would, in turn, mean that $\T$ is well-defined only in a subspace
of $\Pi^{(e)}$. 
But action of $\T$ in eqs.(\ref{Taction}) guarantees that it should be
well-defined on all over $\Pi^{(e)}$ \cite{mukhopadhyay00}. The origin
of this problem is simply the ``wrong'' way of writing $X$ in
eq.(\ref{X}). The last term on the right hand side of eq.(\ref{X})
should be written in such a way that the summation over the tensor indices  
is restricted to the set $\K$ for reason that has been explained in
sec.\ref{s:BS}. 
This immediately leads to,
\beq
|\ev\R = |\mbv, \ev\R ~.
\label{evenfinal}
\eeq
Plugging in the result (\ref{stateadot}) into the third equation of
(\ref{CAe}) one gets,
\bea
|\odd\R &=& - \T \lt[{\sinh \lt(\sqrt{X} \rt) \over \sqrt{X}} 
\lt(\sum_{n>0} S^a_{-n}\St^a_{-n} |\dot a\R \otimes \widetilde{|\dot a
\R} \rt) \rt]
\label{oddfinal}
\eea
Using the Fiertz identity (\ref{fiertzthree}) one can further
manipulate \cite{nemani},
\bea
\sum_{n>0} S^a_{-n}\St^a_{-n} |\dot a\R \otimes \widetilde{|\dot a
\R} = Y^{\dot a \dot b} |\dot a\R \otimes \widetilde{|\dot b\R} 
\eea
where,
\bea
Y^{\dot a \dot b} &=& \sum_{n=1}^{\infty} \lt[ {1\over 8} 
\sum_I \g^I_{a \dot a} \g^I_{b \dot b} S^a_{-n} \St^b_{-n}
+ {1 \over 48} \sum_{IJK} \g^{IJK}_{a\dot a} \g^{IJK}_{b\dot b}
S^a_{-n}  \tilde S^b_{-n} \rt] ~,
\label{Yadotbdot}
\eea
Using this form it is now straightforward to find the action of $\T$ in 
eq.(\ref{oddfinal}). One gets,
\bea
|\odd\R &=&  |\mbv, \odd\R ~,
\eea

\sectiono{Open-Closed Duality}
\label{a:duality}

Here we shall verify the open-closed world-sheet duality by using the 
boundary state (\ref{nonbps}, \ref{nonbpstwo}) and the open string 
spectrum obtained in the BCFT discussion (see
subsecs.\ref{ss:orbifold} and \ref{sss:prescription}). We shall show 
that the duality is satisfied only for the desired value of the 
normalization constant $\nNp$ in eq.(\ref{nonbps}). 

The form of the fermionic part of the boundary state as seen in
eqs.(\ref{nonbpstwo}, \ref{XM}, \ref{YM}) looks quite
complicated. Also the commutation relations among the $J$ oscillators
in eq.(\ref{Jmn}) are not simple enough to result in an easy
computation of the cylinder diagram in closed string channel. However,
using the bosonization and refermionization method of sec.\ref{s:BS}
this
computation can be enormously simplified. This is because of the
simplified form of $|e\R\R$ in eq.(\ref{e-nsns}) and the fact that
the light-cone hamiltonian takes the same form both in terms of the
oscillators of $\tilde S^a(\zb)$ and $\bar S^a(\zb)$.
\bea
H_c &=& {p^2\over 2} + \sum_{n>0,I} \lt( \al^I_{-n} \al^I_n + \tilde
\al^I_{-n} \tilde \al^I_n \rt) +
\sum_{n>0,a} n \lt( S^a_{-n} S^a_n + \tilde S^a_{-n}
\tilde S^a_n \rt)~, \cr
&=& {p^2\over 2} + \sum_{n>0,I} \lt( \al^I_{-n} \al^I_n + \tilde
\al^I_{-n} \tilde \al^I_n \rt) + 
\sum_{n>0,a} n \lt( S^a_{-n} S^a_n + \bar S^a_{-n} \bar S^a_n \rt)~.
\label{Hc}
\eea
The closed string channel expression for the partition function of
open strings ending on the non-BPS D-brane is given by,
\bea
Z &=&  \int_0^{\infty} dl Z(l)~, \cr
Z(l)&=& \langle \nbps ~, p| e^{-2\pi lH_c} |\nbps ~,p\R ~,
\label{Zclosed}
\eea
Standard computations give the following results,
\beq
Z(l) = Z_0(l) . Z_B(l) . Z_F(l)~,
\label{Zclosedtwo}
\eeq
where $Z_0(l)$, $Z_B(l)$ and $Z_F(l)$ are the contributions coming
from the bosonic zero-modes, bosonic oscillators and fermionic
oscillators respectively. Using the inner product 
$\langle k^{\prime}_{\perp}|k_{\perp}\R =
\delta^{(9-p)}(k^{\prime}_{\perp}-k_{\perp})$ and defining
$q=e^{-2\pi l}$ we get,
\beq
Z_0(l) = \lt( \nNp ~\rt)^2 l^{-(9-p)/2}~, \quad Z_B(l) = {q^{2/3} \over
f_1(q)^8}~, \quad Z_F(l) = {1\over 2} q^{-2/3} f_2(q)^8~.
\label{ZoZB}
\eeq
The standard functions $f_i(q)$ that are relevant for our
computations are given by,
\bea
f_1(q) = q^{1/12} \prod_{n=1}^{\infty} (1-q^{2n})~,\quad
f_2(q) = \sqrt{2} q^{1/12} \prod_{n=1}^{\infty} (1+q^{2n}) ~, \quad  
f_4(q) = q^{-1/24}\prod_{n=1}^{\infty} (1-q^{2n-1})~, \cr
\label{fis}
\eea
Defining $t=1/2l$, $\tilde q = e^{-\pi t}$ and using the
transformation properties,
\beq
f_1(e^{-\pi/t}) = \sqrt{t}f_1(e^{-\pi t})~, \quad 
f_2(e^{-\pi/t})= f_4(e^{-\pi t})~,
\label{fitransf}
\eeq
one gets,
\beq
Z=\lt( \nNp ~\rt)^2 2^{(5-p)/2} \int_0^{\infty} dt ~t^{-(3+p)/2}
\lt({f_4(\tilde q)\over f_1(\tilde q)} \rt)^8~.
\label{Zopen}
\eeq
This has to be compared with the open string channel expression,
\beq
Z=\int_0^{\infty} {dt \over 2t} 
\lt[{\rm Tr}_{NS} \lt(e^{-2\pi tH_o}(-1)^F \rt) + 
{\rm Tr}_{R} \lt(e^{-2\pi tH_o}(-1)^F \rt) \rt]~,
\label{Zopen2}
\eeq
where $F$ is the total space-time fermion number of a particular
state. The Hamiltonian is given in the NS and R sectors as,
\bea
H_o = \lt\{ \begin{array}{ll} (\vec{p})^2 + \sum_{n>0,I} \al^I_{-n}\al^I_n
+ \sum_{r>0,a} r S^a_{-r} S^a_r - {1\over 2} ~, & \quad (\hbox{NS})~, \\
& \\
(\vec{p})^2 + \sum_{n>0,I} \al^I_{-n}\al^I_n
+ \sum_{n>0,a} n S^a_{-n} S^a_n~, & \quad (\hbox{R}) ~, \end{array}\rt.
\label{Ho}
\eea
where $\vec p$ is the open string momentum along the world-volume of
the D-brane. Since the R sector spectrum contains equal number of
bosons and fermions at every level we have,
\beq
{\rm Tr}_{R} \lt(e^{-2tH_o}(-1)^F \rt) = 0~.
\label{TrR}
\eeq
The trace in the NS sector gives the following result,
\beq
Z= {V_{p+1}\over (2\pi)^{p+1}} 2^{-(3+p)/2} \int_0^{\infty} dt
~t^{-(3+p)/2} \lt( {f_4(\tilde q) \over f_1(\tilde q)}\rt)^8~,
\label{Zopen3}
\eeq
where we have used $\langle \vec p |\vec p\R = \delta^{(p+1)}(0) =
V_{p+1}/(2\pi)^{p+1}$, $V_{p+1}$ being the (infinite) volume of the $(p+1)$
dimensional world-volume. The two results (\ref{Zclosedtwo}) and
(\ref{Zopen3}) match only for,
\beq
\lt(\nNp ~\rt)^2 = {V_{p+1} \over 16 (2\pi)^{p+1}}~.
\label{nNpsquare}
\eeq
We shall now argue that the above equation implies Sen's
result \cite{sen9805, senrev} which relates
${\nTp}$, the tension of a non-BPS D-brane in type IIB and $\T_p$, the
tension of the corresponding BPS D-brane in type IIA,
\beq
\nTp = \sqrt{2} ~\T_p~.
\label{nTpTp}
\eeq
Boundary state for the BPS D-brane in type IIA which has the same
configuration as the non-BPS D-brane (\ref{nonbps}) in type IIB is given by,
\bea
|\bps,\eta\R_{(IIA)} &=& {\cal N}_p \int d^{(9-p)}k_{\perp} 
\exp\lt( \sum_{n>0} {1\over n} \al^I_{-n} \mbv_{IJ} \tilde \al^J_{-n} 
-i\eta \sum_{n>0} S^a_{-n} \mbs_{a\dot b} \tilde S^{\dot b}_{-n} \rt)
\cr
&& \lt[ \mbv_{IJ} |I\R \otimes \widetilde{|J\R} 
- i \eta \mbc_{\dot a b} |\dot a\R \otimes \widetilde{|b\R }\rt]
\otimes |k_{\perp}\R ~.
\label{bpseta}
\eea
A computation similar to the one performed above shows that the
world-sheet duality is satisfied for the following value
of the normalization constant.
\beq
{\cal N}_p^2 = {V_{p+1}\over 32(2\pi)^2}~.
\label{Npsquare}
\eeq
Now noticing that $\nNp$ and ${\cal N}_p$ are the strengths of the
massless NSNS state 
\beq
\int d^{(9-p)}k_{\perp} \mbv_{IJ} 
|I\R \otimes \widetilde{|J\R} \otimes |k_{\perp}\R~, 
\label{masslessnsns}
\eeq
in the boundary states $|\nbps~\R$ and $|\bps,\eta\R_{(IIA)}$ 
respectively, eqs.(\ref{nNpsquare}) and (\ref{Npsquare}) imply (\ref{nTpTp}).
 
\sectiono{Open String Boundary Condition}
\label{a:bc}

Here we shall derive the open string boundary condition
(\ref{Sacondition}) from a certain quadratic gluing condition
satisfied by the covariant expression (\ref{nonbpstwo}). Although 
the set of oscillators $J_{mn}$, $J^{IJ}_{mn}$ and $J^{IJKL}_{mn}$
do not have simple commutation relations among themselves and finding
such a gluing condition seems to be difficult, one can actually simplify
this task by using the bosonization and refermionization trick of
sec.\ref{s:BS}. Since $|\hbox{BPS}, -1,\eta \R_{\bar S}$ takes the
form of the fermionic part of a BPS D-instanton boundary state with
the replacement indicated below eq.(\ref{e-nsns}), it satisfies the 
gluing condition
(\ref{bpsgluing}) with the same replacement and with $\ms_{ab} =
\dt_{ab}$. Using this gluing condition and eq.(\ref{e-nsns}) one
readily finds \cite{nemani},
\bea
S^a_m S^b_n |e\R\R = \bar S^b_{-n} \bar S^a_{-m} |e\R \R ~, \quad \forall
m,n \in \ZZZ~.
\label{e-gluing-modes}
\eea
This can be re-written in terms of the local fields on the cylinder in
the closed string channel as, 
\bea
\lt[S^a(0,\sigma) S^b(0,\sigma^{\prime}) - \bar S^b(0,\sigma^{\prime})
\bar S^a(0,\sigma)  \rt] |e\R\R =0~,
\label{e-gluing-fields}
\eea
where the first argument is the value of the world-sheet time $\tau$
in closed string channel. Using the standard method one can go
to the open string channel to get the following open string boundary
condition on the upper half plane,
\bea
S^a(z) S^b(w) + \bar S^b(\wb) \bar S^a(\zb) = 0~, \quad \hbox{at }
z=\zb, ~w=\wb~.
\label{open-bc1}
\eea
Notice that for $z\neq w$ one can write,
\bea
S^a(z) S^b(w) = \bar S^a(\zb) \bar S^b(\wb) ~, \quad \hbox{at }
z=\zb, ~w=\wb~.
\label{open-bc2}
\eea
One can also argue that this equation is correct even when $z=w$
if we take the operators to be normal ordered. We can now rewrite the
right hand side in terms of $\St^a(\zb)$:
\bea
\bar S^a(\zb) \bar S^b(\wb) &=& {1\over 8} \dt_{ab} \lt(\bar S^c(\zb)
\bar S^c(\wb)\rt) + {1\over 16} \sum_{IJ} \g^{IJ}_{ab}
\lt(\bar S^c(\zb) \g^{IJ}_{cd} \bar S^d(\wb) \rt) + \cr &&
{1\over 384} \sum_{IJKL} \g^{IJKL}_{ab}
\lt(\bar S^c(\zb) \g^{IJKL}_{cd} \bar S^d(\wb) \rt) ~, \cr
&=& {1\over 8} \dt_{ab} \lt(\St^c(\zb) \St^c(\wb)\rt) +
{1\over 16} \sum_{IJ} \lamb_{\{IJ\}} \g^{IJ}_{ab}
\lt(\St^c(\zb) \g^{IJ}_{cd} \St^d(\wb) \rt) + \cr &&
{2\over 384} \sum_{\{IJKL\}\in \K} \lamb_{\{IJKL\}} \g^{IJKL}_{ab}
\lt(\St^c(\zb) \g^{IJKL}_{cd} \St^d(\wb) \rt) ~.
\label{chi-St}
\eea
We have used the Fiertz identity (\ref{fiertzone}) in the first step
where all the tensor indices are summed over freely. In the second
step we have performed the bosonization and refermionization trick of
sec.\ref{s:BS} in the reverse direction. Finally, using the result
(\ref{chi-St}) in eq.(\ref{open-bc2}) one arrives at the condition
(\ref{Sacondition}).

\sectiono{One-Point Functions}
\label{a:onepoint}

Here we shall verify the relation (\ref{one-point}) with $C=1$ for the
states $|\Phi^{(IJK),(MNP)}_{m,n}\R$, $|\Phi^{IJ}_{m,n,p,q}\R$
$|\Phi^{(MN),(M^{\prime}N^{\prime}),IJ}_{m,n,p,q}\R$ and
$|\Phi^{(MNPQ),(M^{\prime}N^{\prime}P^{\prime}Q^{\prime}),IJ}_{m,n,p,q}\R$
as defined in eqs.(\ref{states}). The corresponding vertex operators
are given by,
\bea
\Phi^{(IJK),(MNP)}_{m,n}(\ze,\zeb) &=& - \g^{IJK}_{a\dot a}
\g^{MNP}_{b\dot b} \Phi^{a b \dot a \dot b}_{m,n}(\ze,\zeb) ~, \cr &&
\cr
\Phi^{IJ}_{m,n,p,q}(\ze,\zeb) &=&
\Phi^{aabbIJ}_{m,n,p,q}(\ze,\zeb)~,\cr && \cr
\Phi^{(MN),(M^{\prime}N^{\prime}),IJ}_{m,n,p,q}(\ze,\zeb) &=&
\g^{MN}_{ab} \g^{M^{\prime}N^{\prime}}_{cd}
\Phi^{abcdIJ}_{m,n,p,q}(\ze,\zeb)~, \cr && \cr
\Phi^{(MNPQ),(M^{\prime}N^{\prime}P^{\prime}Q^{\prime}),IJ}_{m,n,p,q}
(\ze,\zeb)&=& \g^{MNPQ}_{ab}
\g^{M^{\prime}N^{\prime}P^{\prime}Q^{\prime}}_{cd}
\Phi^{abcdIJ}_{m,n,p,q}(\ze,\zeb)~,
\label{tensor-vertices}
\eea
respectively. The operator $\Phi^{ab \dot a \dot b}_{m,n}(\ze,\zeb)$
is defined in eq.(\ref{spinor-vertex1}) and,
\bea
\Phi^{abcdIJ}_{m,n,p,q}(\ze,\zeb) &=&
\oint_{\ze} {d\ze_1 d\ze_2\over (2\pi i)^2 }
\oint_{\zeb} {d\zeb_3 d\zeb_4 \over (2\pi i)^2}
(\ze_1-\ze)^{-m-1/2} (\ze_2-\ze)^{-n-1/2} \cr &&
(\zeb_3-\zeb)^{-p-1/2} (\zeb_4-\zeb)^{-q-1/2}
S^a(\ze_1) S^b(\ze_2) \St^c(\zeb_3) \St^d(\zeb_4) \psi^I(\ze)
\psit^J(\zeb) ~. \cr &&
\label{spinor-vertices}
\eea
The non-trivial inner products can be computed to be,
\bea
\langle \Phi^{(IJK),(MNP)}_{m,n}|
\Phi^{(I^{\prime}J^{\prime}K^{\prime}),(M^{\prime}N^{\prime}P^{\prime})
}_{m^{\prime},n^{\prime}} \R &=& 64 \dt_{m,m^{\prime}}
~\dt_{n,n^{\prime}}~\D_{(IJK),(I^{\prime}J^{\prime}K^{\prime})}
\D_{(MNP),(M^{\prime}N^{\prime}P^{\prime})}~, \cr && \cr
\langle \Phi^{IJ}_{m,n,p,q}|
\Phi^{I^{\prime}J^{\p}}_{m^{\p},n^{\p},p^{\p},q^{\p}} \R &=&
64 (\dt_{m,m^{\p}}~\dt_{n,n^{\p}} - \dt_{m,n^{\p}}~\dt_{n,m^{\p}}) \cr &&
(\dt_{p,p^{\p}}~\dt_{q,q^{\p}} - \dt_{p,q^{\p}}~\dt_{q,p^{\p}})
\dt_{I,I^{\p}}~ \dt_{J,J^{\p}}~, \cr && \cr
\langle \Phi^{(KL),(MN),IJ}_{m,n,p,q} |
\Phi^{(K^{\p}L^{\p}),
(M^{\p}N^{\p}),I^{\p}J^{\p}}_{m^{\p},n^{\p},p^{\p},q^{\p}}\R &=&
64 (\dt_{m,m^{\p}}~\dt_{n,n^{\p}} + \dt_{m,n^{\p}}~\dt_{n,m^{\p}}) \cr &&
(\dt_{p,p^{\p}}~\dt_{q,q^{\p}} + \dt_{p,q^{\p}}~\dt_{q,p^{\p}})
\dt_{I,I^{\p}} ~ \dt_{J,J^{\p}} \cr &&
\D_{(KL),(K^{\p}L^{\p})} \D_{(MN),(M^{\p}N^{\p})} ~, \cr && \cr
\langle \Phi^{(KLMN),(PQRS),IJ}_{m,n,p,q} |
\Phi^{(K^{\p}L^{\p}M^{\p}N^{\p}),
(P^{\p}Q^{\p}R^{\p}S^{\p}),I^{\p}J^{\p}}_{m^{\p},n^{\p},p^{\p},q^{\p}}\R &=&
64 (\dt_{m,m^{\p}}~\dt_{n,n^{\p}} - \dt_{m,n^{\p}}~\dt_{n,m^{\p}}) \cr &&
(\dt_{p,p^{\p}}~\dt_{q,q^{\p}} - \dt_{p,q^{\p}}~\dt_{q,p^{\p}})
\dt_{I,I^{\p}} ~ \dt_{J,J^{\p}} \cr &&
\D_{(KLMN),(K^{\p}L^{\p}M^{\p}N^{\p})}
\D_{(PQRS),(P^{\p}Q^{\p}R^{\p}S^{\p})} ~, \cr
&& \hbox{for } \{K,L,M,N\},\{K^{\p},L^{\p},M^{\p},N^{\p}\}, \cr &&
\{P,Q,R,S\},\{P^{\p},Q^{\p},R^{\p},S^{\p}\} \in \K~.
\label{inner-products}
\eea
where we have introduced,
\bea
\D_{(I_1 \cdots I_n),(I_1^{\prime}\cdots I^{\p}_n)} \equiv
\sum_{{\cal P}} \hbox{sign} {\cal P} ~ \dt_{\{I_1\cdots I_n\},
{\cal P}\{I_1^{\prime}\cdots I_n^{\p}\}} ~.
\label{Delta}
\eea
The summation goes over $n!$ terms. $\{I_1,\cdots I_n\}$ is an ordered
set while ${\cal P} \{I_1,\cdots I_n\}$ is another ordered set obtained by
applying the permutation ${\cal P}$ on $\{I_1,\cdots I_n\}$. $\hbox{sign}{\cal P}$
is $1$ if ${\cal P}$ is even and $(-1)$ otherwise.
$\dt_{\{I_1\cdots I_n\},\{I_1^{\prime}\cdots I_n^{\p}\}}$ is the
$n$-dimensional Kronecker delta function which is $1$ only when the
two ordered sets $\{I_1\cdots I_n\}$ and $\{I_1^{\prime}\cdots
I_n^{\p}\}$ are equal and zero otherwise. Using the above inner products
and eq.(\ref{nonbpstwo}) one gets the following results,
\bea
\langle \langle e |\Phi^{(IJK),(MNP)}_{m,n} \R &=& 8 \dt_{m,n} ~
\lamb_{\{IJK\}} \D_{(IJK),(MNP)}~, \label{bstate-result1} \\
&& \cr
\langle \langle e |\Phi^{IJ}_{m,n,p,q}\R &=& 8 (\dt_{m,p}~\dt_{n,q} -
\dt_{m,q}~\dt_{n,p}) \lamb_I \delta_{IJ} ~, \label{bstate-result2} \\
&& \cr
\langle \langle e |\Phi^{(MN),(PQ),IJ}_{m,n,p,q} \R &=&
8 (\dt_{m,p}~\dt_{n,q} + \dt_{m,q}~\dt_{n,p}) \lamb_I \lamb_{\{MN\}}
\delta_{IJ} \D_{(MN),(PQ)}~, \label{bstate-result3} \\
&& \cr
\langle \langle e |\Phi^{(KLMN),(PQRS),IJ}_{m,n,p,q}\R &=&
8(\dt_{m,p}~\dt_{n,q} - \dt_{m,q}~\dt_{n,p})
\lamb_I \lamb_{\{KLMN\}} \delta_{IJ}
\D_{(KLMN),(PQRS)} ~, \cr &&
\hbox{for } \{K,L,M,N\}, \{P,Q,R,S\} \in \K~.
\label{bstate-result4}
\eea
By going through an analysis similar to the one done in subsec.
\ref{sss:onepoint} one can
write the one-point functions of the vertices in
(\ref{tensor-vertices}) in terms of certain correlation functions on
the full plane.
\bea
\langle \Phi^{(IJK),(MNP)}_{m,n}(0,0) \R_D &=&
\lamb_{\{MNP\}}
\oint_i {dz \over 2\pi i}
\oint_{-i} {dw \over 2\pi i} {\cal J}_{m,n}(z, w) \cr
&& \g^{IJK}_{a\dot a} \g^{MNP}_{b\dot b}
\langle \CS^a(z) \CS^b(w) \CS^{\dot a}(i) \CS^{\dot  b}(-i) \R~,
\label{disk1} \\
&& \cr
\langle \Phi^{IJ}_{m,n,p,q}(0,0) \R_D &=&
\lamb_J \oint_i {dz_1 dw_1 \over (2\pi i)^2}
\oint_{-i} {dz_2 dw_2 \over (2\pi i)^2}
{\cal J}_{m,n,p,q}(z_1,w_1,z_2,w_2) \cr
&& \langle \CS^a(z_1) \CS^a(w_1) \CS^c(z_2) \CS^c(w_2) \Psi^I(i)
\Psi^J(-i)\R~, \label{disk2} \\
&& \cr
\langle \Phi^{(MN),(PQ),IJ}_{m,n,p,q}(0,0) \R_D &=&
\lamb_J \lamb_{\{PQ\}}
\oint_i {dz_1 dw_1 \over (2\pi i)^2}
\oint_{-i} {dz_2 dw_2 \over (2\pi i)^2}
{\cal J}_{m,n,p,q}(z_1,w_1,z_2,w_2) \cr
&& \g^{MN}_{ab} \g^{PQ}_{cd}
\langle \CS^a(z_1) \CS^b(w_1) \CS^c(z_2) \CS^d(w_2) \Psi^I(i)
\Psi^J(-i)\R~, \cr && \label{disk3} \\
&& \cr
\langle \Phi^{(KLMN),(PQRS),IJ}_{m,n,p,q}(0,0) \R_D &=&
\lamb_J \lamb_{\{PQRS\}}
\oint_i {dz_1 dw_1 \over (2\pi i)^2}
\oint_{-i} {dz_2 dw_2 \over (2\pi i)^2}
{\cal J}_{m,n,p,q}(z_1,w_1,z_2,w_2) \cr
&& \g^{KLMN}_{ab} \g^{PQRS}_{cd}
\langle \CS^a(z_1) \CS^b(w_1) \CS^c(z_2) \CS^d(w_2) \Psi^I(i)
\Psi^J(-i)\R~,\cr &&
\label{disk4}
\eea
where ${\cal J}_{m,n}(z,w)$ is given in eq.(\ref{Jacobian-mn}) and,
\bea
{\cal J}_{m,n,p,q}(z_1,w_1,z_2,w_2) &=& 8 (z_1-i)^{-m-1/2}
(z_1+i)^{m-1/2} (w_1-i)^{-n-1/2} (w_1+i)^{n-1/2} \cr &&
(z_2+i)^{-p-1/2} (z_2-i)^{p-1/2} (w_2+i)^{-q-1/2} (w_2-i)^{q-1/2}~.
\cr &&
\label{Jacobian-mnpq}
\eea
The correlation functions appearing in the above integrands have been
computed in appendix \ref{a:correlators} in terms of certain
functions. The full set of results of the required integrals involving
these functions have been listed in appendix \ref{a:integrals}. To
compute the right hand sides of eqs. (\ref{disk1}), (\ref{disk2}),
(\ref{disk3}) and (\ref{disk4}) one needs to use eqs.
(\ref{int2}), (\ref{int3}), (\ref{int4}, \ref{int5}, \ref{int6}) and
(\ref{int7}, \ref{int8}, \ref{int9}) respectively. The final results 
show equality
of the one-point functions (\ref{disk1}), (\ref{disk2}), (\ref{disk3}),
(\ref{disk4}) with the boundary state results (\ref{bstate-result1}),
(\ref{bstate-result2}), (\ref{bstate-result3}), (\ref{bstate-result4})
respectively.

\sectiono{Correlation Functions on the Plane}
\label{a:correlators}

Certain 4 and 6-point correlation functions on the full plane are
needed to compute the closed string one-point functions
(\ref{disk0}), (\ref{disk1}), (\ref{disk2}), (\ref{disk3}) and
(\ref{disk4}). The index structures and various symmetry properties
allow us to write down these correlation functions in the following
forms,
\bea
\g^I_{a \dot a} \g^J_{b \dot b}
\langle \CS^a(z) \CS^b(w) \CS^{\dot a}(u) \CS^{\dot b}(v)\R &=&
\dt_{IJ} \Ft^{(1,1)}(z,w,u,v)~, \label{corr0}\\ && \cr
\g^{IJK}_{a \dot a} \g^{MNP}_{b \dot b}
\langle \CS^a(z) \CS^b(w) \CS^{\dot a}(u) \CS^{\dot b}(v)\R
&=& \D_{(IJK),(MNP)} \Ft^{(3,3)}(z,w,u,v)~, \cr && \label{corr1} \\
\langle \CS^a(z_1) \CS^a(w_1) \CS^c(z_2) \CS^c(w_2)\Psi^I(u)
\Psi^J(v) \R &=& \dt_{IJ} F^{(1,1)}(z_1,w_1,z_2,w_2,u,v)~, \cr &&
\label{corr2}
\eea
\bea
&& \displaystyle{\g^{MN}_{ab} \g^{PQ}_{cd}
\langle \CS^a(z_1) \CS^b(w_1) \CS^c(z_2) \CS^d(w_2)\Psi^I(u)
\Psi^J(v) \R } \cr
&=& \dt_{IJ} \D_{(MN),(PQ)}
F_1^{(3,3)}(z_1,w_1,z_2,w_2,u,v) + \cr
&&\lt( \dt_{QJ} \D_{(MN),(IP)} + \dt_{PJ} \D_{(MN),(QI)} \rt)
F_2^{(3,3)}(z_1,w_1,z_2,w_2,u,v)   - \cr
&&\lt( \dt_{QI} \D_{(MN),(JP)} + \dt_{PI} \D_{(MN),(QJ)} \rt)
F_2^{(3,3)}(z_1,w_1,z_2,w_2,v,u)  ~,  \label{corr3} \\ && \cr
&& \displaystyle{\g^{KLMN}_{ab} \g^{PQRS}_{cd}
\langle \CS^a(z_1) \CS^b(w_1) \CS^c(z_2) \CS^d(w_2)\Psi^I(u) \Psi^J(v) \R} \cr
&=& \dt_{IJ} \lt(\D_{(KLMN),(PQRS)} + \eps^{KLMNPQRS} \rt)
F^{(5,5)}_1(z_1,w_1,z_2,w_2,u,v) + \cr
&&D_{(KLMN),(PQRS),IJ} F^{(5,5)}_2(z_1,w_1,z_2,w_2,u,v) -
D_{(KLMN),(PQRS),JI} F^{(5,5)}_2(z_1,w_1,z_2,w_2,v,u)~, \cr &&
\label{corr4}
\eea
where the notation $\Del_{(IJ\cdots)(KL\cdots)}$ has been introduced
in eq.(\ref{Delta}) and
\bea
D_{(KLMN),(PQRS),IJ} &=& \dt_{KI} \lt(\dt_{PJ}\D_{(LMN),(QRS)}
-\dt_{QJ}\D_{(LMN),(PRS)} \rt. \cr
&& \lt. -\dt_{RJ}\D_{(LMN),(QPS)}-\dt_{SJ}\D_{(LMN),(QRP)} \rt) \cr &&
-\dt_{LI} \lt(\dt_{PJ}\D_{(KMN),(QRS)}-\dt_{QJ}\D_{(KMN),(PRS)} \rt. \cr
&& \lt. -\dt_{RJ}\D_{(KMN),(QPS)}-\dt_{SJ}\D_{(KMN),(QRP)} \rt) \cr &&
- \dt_{MI}\lt( \dt_{PJ}\D_{(LKN),(QRS)} -\dt_{QJ}\D_{(LKN),(PRS)} \rt. \cr
&& \lt. -\dt_{RJ}\D_{(LKN),(QPS)}-\dt_{SJ}\D_{(LKN),(QRP)} \rt) \cr &&
- \dt_{NI} \lt(\dt_{PJ}\D_{(LMK),(QRS)} -\dt_{QJ}\D_{(LMK),(PRS)}\rt. \cr
&& \lt. -\dt_{RJ}\D_{(LMK),(QPS)}-\dt_{SJ}\D_{(LMK),(QRP)}  \rt)~.
\label{def-D}
\eea
The various functions have the following symmetry properties,
\bea
\Ft^{(1,1)}(z,w,u,v) &=&  \Ft^{(1,1)}(w,z,v,u)~,\label{Ftld11-symm} \\
\Ft^{(3,3)}(z,w,u,v) &=&  \Ft^{(3,3)}(w,z,v,u)~,\label{Ftld33-symm} \\
F^{(1,1)}(z_1,w_1,z_2,w_2,u,v)
&=& - F^{(1,1)}(w_1,z_1,z_2,w_2,u,v)~,\cr
&=& -F^{(1,1)}(z_1,w_1,w_2,z_2,u,v)~, \cr
&=& -F^{(1,1)}(z_1,w_1,z_2,w_2,v,u)~, \cr
&=& F^{(1,1)}(z_2,w_2,z_1,w_1,u,v) ~, \label{F11-symm} \\
&& \cr
F_1^{(3,3)}(z_1,w_1,z_2,w_2,u,v)
&=& F_1^{(3,3)}(w_1,z_1,z_2,w_2,u,v)~,\cr
&=& F_1^{(3,3)}(z_1,w_1,w_2,z_2,u,v)~,\cr
&=& - F_1^{(3,3)}(z_1,w_1,z_2,w_2,v,u)~,\cr
&=& F_1^{(3,3)}(z_2,w_2,z_1,w_1,u,v)~, \label{F331-symm}\\
&& \cr
F_2^{(3,3)}(z_1,w_1,z_2,w_2,u,v)
&=& F_2^{(3,3)}(w_1,z_1,z_2,w_2,u,v)~, \cr
&=& F_2^{(3,3)}(z_1,w_1,w_2,z_2,u,v)~, \label{F332-symm}\\
&& \cr
F_1^{(5,5)}(z_1,w_1,z_2,w_2,u,v)
&=&-F_1^{(5,5)}(w_1,z_1,z_2,w_2,u,v) ~, \cr
&=&-F_1^{(5,5)}(z_1,w_1,w_2,z_2,u,v) ~, \cr
&=&-F_1^{(5,5)}(z_1,w_1,z_2,w_2,v,u)~, \cr
&=& F_1^{(5,5)}(z_2,w_2,z_1,w_1,u,v)~, \\
\label{F55a-symm}
&& \cr
F_2^{(5,5)}(z_1,w_1,z_2,w_2,u,v)
&=& - F_2^{(5,5)}(w_1,z_1,z_2,w_2,u,v) ~, \cr
&=& - F_2^{(5,5)}(z_1,w_1,w_2,z_2,u,v) ~, \cr
&=& - F_2^{(5,5)}(z_2,w_2,z_1,w_1,v,u)~.
\label{F552-symm}
\eea
To compute the above functions we bosonize the holomorphic fermions
$\CS^a(z)$, $\CS^{\dot a}(z)$ and $\Psi^I(z)$ and use
Mathematica. This method requires a
full set of consistent formulas for the basis vectors for the various
weight lattices, cocycle factors and the corresponding gamma matrix
representation. We take the sets of basis vectors $e_i$, $\dt_i$ and
$\bar \dt_i$ ($i=1,\cdots ,4$) for the vector, spinor and conjugate
spinor weight lattices respectively to be as given in
\cite{mukhopadhyay00}. For cocycle factors and the corresponding gamma
matrices we follow the convention of \cite{kostelecky87}. The explicit
formulas can be found in appendix \ref{a:convention}. We give the final
results below. The functions involved in the four point correlators
are computed to be,
\bea
&&
\begin{array}{l}
\Ft^{(1,1)}(z,w,u,v) = - {24 i \over \sqrt{(z-u)(z-v)(w-u)(w-v)}}
+ {8i \sqrt{(z-v)(w-u)}\over (u-v)(z-w) \sqrt{(z-u)(w-v)}} ~,
\end{array}\\
\label{Ftld11}
&&
\begin{array}{l}
\Ft^{(3,3)}(z,w,u,v) = {8 i \sqrt{(z-u)(w-v)} \over (u-v)(z-w)
\sqrt{(z-v)(w-u)}} ~,
\end{array}
\label{Ftld33}
\eea
The functions involved in the six-point correlators are given by,
\bea
F^{(1,1)}(z_1,w_1,z_2,w_2,u,v) &=& F_1(z_1,w_1,z_2,w_2,u,v) -
F_1(w_1,z_1,z_2,w_2,u,v) -\cr &&
F_1(z_1,w_1,w_2,z_2,u,v) - F_1(z_1,w_1,z_2,w_2,v,u) + \cr &&
F_2(z_1,w_1,z_2,w_2,u,v) - F_2(w_1,z_1,z_2,w_2,u,v) -\cr &&
F_2(z_1,w_1,w_2,z_2,u,v) - F_2(z_1,w_1,z_2,w_2,v,u) ~,  \cr &&
\label{F11} \\
F_1^{(3,3)}(z_1,w_1,z_2,w_2,u,v) &=& {1\over 3}F_1(z_1,w_2,z_2,w_1,u,v) +
{1\over 3}F_1(w_1,w_2,z_2,z_1,u,v) + \cr &&
{1\over 3}F_1(z_1,z_2,w_2,w_1,u,v) -
{1\over 3}F_1(z_1,w_2,z_2,w_1,v,u) ~, \cr &&
\label{F331} \\
F_2^{(3,3)}(z_1,w_1,z_2,w_2,u,v) &=& F_3(z_1,w_1,z_2,w_2,u,v) +
F_3(w_1,z_1,z_2,w_2,u,v) + \cr &&
F_3(z_1,w_1,w_2,z_2,u,v) + F_3(w_1,z_1,w_2,z_2,u,v) ~, \cr &&
\label{F332} \\
F_1^{(5,5)}(z_1,w_1,z_2,w_2,u,v) &=& - {1\over 3}
F_1(z_1,w_2,z_2,w_1,u,v) + {1 \over 3} F_1(w_1,w_2,z_2,z_1,u,v) \cr
&& +{1\over 3} F_1(z_1,w_2,z_2,w_1,v,u) + {1\over 3}
F_1(z_1,z_2,w_2,w_1,u,v) ~, \cr &&
\label{F551} \\
F_2^{(5,5)}(z_1,w_1,z_2,w_2,u,v) &=& - F_2(z_1,w_2,z_2,w_1,u,v)
- F_2(z_1,w_2,z_2,w_1,v,u) ~,
\label{F552}
\eea
where
\bea
\begin{array}{l}
F_1(z_1,w_1,z_2,w_2,u,v)
= {12 \sqrt{(z_1-u)(z_2-u)(w_1-v)(w_2-v)}\over (u-v)(z_1-w_1)(z_2-w_2)
\sqrt{(z_1-v)(z_2-v)(w_1-u)(w_2-u)} } ~, \\ \\
F_2(z_1,w_1,z_2,w_2,u,v)
= -{4(z_1-w_2)(z_2-w_1) \sqrt{(z_1-v)(z_2-u)(w_1-u)(w_2-v)} \over
(z_1-z_2)(w_1-w_2)(u-v)(z_1-w_1)(z_2-w_2)
 \sqrt{(z_1-u)(z_2-v)(w_1-v)(w_2-u)}}~, \\ \\
F_3(z_1,w_1,z_2,w_2,u,v)
= -{4\sqrt{(w_1-v)(w_2-u)} \over
 (w_1-w_2) \sqrt{(z_1-u)(z_1-v)(w_1-u)(w_2-v) (z_2-u)(z_2-v)}}~,
\end{array} \cr
\label{F1F2F3}
\eea

\sectiono{Bosonization and Gamma Matrices}
\label{a:convention}

Here we present the basic formulas for the bosonization of the
holomorphic fields $\CS^a(z)$, $\CS^{\dot a}(z)$, and $\Psi^I(z)$
on the full complex plane and the corresponding gamma matrix representation.
We take the sets of four dimensional basis vectors $e_i$, $\dt_i$ and
$\bar \dt_i$ ($i=1,\cdots ,4$) for the vector, spinor and conjugate
spinor weight lattices respectively to be as given in
\cite{mukhopadhyay00},
\bea
\begin{array}{llll}
e_1=(1,0,0,0),& e_2=(0,1,0,0),& e_3=(0,0,1,0),& e_4=(0,0,0,1)~, \\
\delta_1 ={1\over 2}(-1,1,-1,1),& \delta_2 = {1\over 2}(-1,1,1,-1),&
\delta_3= {1\over 2}(1,1,1,1),& \delta_4= {1\over 2}(1,1,-1,-1)~, \\
\delb_1= {1\over 2}(1,1,-1,1),&\delb_2={1\over 2}(1,1,1,-1) ,
&\delb_3={1\over 2}(-1,1,1,1) ,&\delb_4={1\over 2}(-1,1,-1,-1) ~,
\end{array} \cr
\label{weights}
\eea
The spinors corresponding to the above weight vectors are defined to
be,
\bea
\lambda^{\pm e_j}(z) &=& {1\over \sqrt{2}} \lt(\Psi^{\mu=2j-1}(z) \mp
i \Psi^{\mu=2j}(z) \rt)~,\cr
\chi^{\pm \delta_j}(z) &=& \exp \lt(-i {\pi\over 2} \delta_j.M_4\delta_j
\rt){1\over \sqrt{2}} \lt(\CS^{a=2j-1}(z) \mp
i \CS^{a=2j}(z) \rt)  ~,\cr
\xi^{\pm \delb_j}(z) &=& \exp \lt(-i {\pi\over 2} \bar \delta_j.M_4\bar \delta_j
\rt)
{1\over \sqrt{2}} \lt(\CS^{\dot a=2j-1}(z) \mp
i \CS^{\dot a=2j}(z) \rt)~,
\label{weightspinor}
\eea
where,
\bea
M_4 = \pmatrix{0&0&0&0 \cr 1&0&0&0 \cr 1&1&0&0 \cr -1&1&-1&0}~.
\label{M4}
\eea
The extra phases in the last two equations in (\ref{weightspinor})
will be explained later. The bosonization is given by,
\bea
\lambda^{\pm e_j}(z) &=& \exp \lt( \pm i e_j.\phi \rt)(z)
C_{\pm  e_j}(\hat p)~, \cr
\chi^{\pm \delta_j}(z) &=& \exp \lt( \pm i \delta_j.\phi \rt)(z)
C_{\pm \delta_j}(\hat p) ~, \cr
\xi^{\pm \delb_j}(z) &=& \exp \lt( \pm i \delb_j.\phi \rt)(z)
C_{\pm \delb_j}(\hat p) ~,
\label{bosonization}
\eea
where $\phi(z)$ is a holomorphic scalar field normalized such that
$\phi(z) \phi(w) \sim - \ln(z-w)$. Following \cite{kostelecky87} we
take the cocycle factor $C_w(\hat p)$ corresponding to the weight $w$
to be,
\bea
C_w(\hat p) &=& \exp\lt(i \pi w.M_4 \hat p \rt)~,
\label{cocycle}
\eea
where $\hat p$ is the momentum conjugate to the zero mode of $\phi(z)$.
Defining the vector, spinor and conjugate spinor indices $\hat I$,
$A$ and $\dot A =1,\cdots 8$ corresponding to the lattice basis such
that,
\bea
&&
E_{\hat I} = \lt\{\begin{array}{ll}
e_j & \hbox{ for } \hat I = 2j-1  \\
-e_j &\hbox{ for } \hat I = 2j
\end{array} \rt. ~, \quad
\Del_A = \lt\{\begin{array}{ll}
\delta_j  & \hbox{ for } A= 2j-1  \\
-\delta_j & \hbox{ for } A= 2j
\end{array} \rt. ~, \cr
&&
\Delb_{\dot A} = \lt\{\begin{array}{ll}
\delb_j  & \hbox{ for } \dot A = 2j-1 \\
-\delb_j & \hbox{ for } \dot A = 2j
\end{array} \rt. ~, \quad j=1,\cdots , 4~,
\label{lat-indices}
\eea
the corresponding gamma matrices \cite{kostelecky87}\footnote{The
gamma matrix representation taken in \cite{mukhopadhyay00} is
consistent with this choice of the basis vectors on the weight
lattices, but corresponds to different choice of cocycle factors which
were not needed for the computations considered there.} are given by,
\bea
(\g^{\hat I})^A_{~\dot A} &=& \sqrt{2}
\exp \lt(i\pi E_{\hat I}.M_4\Delb_{\dot A} \rt) \delta_{E_{\hat I}+\Del_A,
\Delb_{\dot A}}~, \cr
(\gb^{\hat I})^{\dot A}_{~A} &=& \sqrt{2}
\exp\lt(i\pi E_{\hat I}.M_4\Del_A \rt) \delta_{E_{\hat I}+\Delb_{\dot A},
\Del_A}~.
\label{lat-gamma}
\eea
The spinor indices are raised and lowered by the charge conjugation
matrices,
\bea
C^{AB} &=& \exp \lt(-i\pi \Del_A.M_4 \Del_A \rt) \delta_{\Del_A ,
-\Del_B}~, \cr
C^{\dot A\dot B} &=& i \exp \lt(-i\pi \Delb_{\dot A}.M_4 \Delb_{\dot
A} \rt) \delta_{\Delb_{\dot A} ,-\Delb_{\dot B}}~.
\label{lat-C}
\eea
The phases in the last two equations in (\ref{bosonization})
correspond to one choice such that the charge conjugation matrices are
given by the identity matrix in the covariant basis, i.e.
$\CS^a(z) \CS^b(z) \sim {\dt^{ab}\over (z-w)}$,
$\CS^{\dot a}(z) \CS^{\dot b}(z) \sim {\dt^{\dot a\dot b}\over
(z-w)}$. In this basis the 16-dimensional gamma matrices are
hermitian: $\lt( \Gamma^I\rt)^{\dagger} =\Gamma^I$. Therefore,
\bea
\gb^I = \lt( \g^I \rt)^{\dagger}~.
\label{hermiticity}
\eea
It turns out that any non-zero element of a $\g^I$ is one of the
following three phases: $e^{i\pi/4},~\pm i e^{-i\pi/4}$. This implies
the following reality property,
\bea
\g^I_{a\dot a} \lt( \g^J_{b\dot b}\rt)^* = \hbox{ real}~.
\label{reality}
\eea
Using the above reality and hermiticity properties one establishes
that $\g^{IJ}$ and $\gb^{IJ}$ are real antisymmetric while
$\g^{IJKL}$ and $\gb^{IJKL}$ are real symmetric.

\vspace{10pt}
\centerline{\it Representation independent relations}
\vspace{10pt}

\noindent
Various Fiertz identities, which we have used very crucially in our
computations, are given by,
\bea
\lambda_1^a \lambda_2^b &=& {1 \over 8} (\lambda_1^c
\lambda_2^c) \, \delta_{ab} + {1 \over 16} (\lambda_1^c
\gamma^{IJ}_{cd} \lambda_2^d ) \, \gamma^{IJ}_{ab} + {1 \over 384}
( \lambda_1^c \gamma^{IJKL}_{cd} \lambda_2^d ) \,
\gamma^{IJKL}_{ab}, \label{fiertzone} \\
\lambda_1^{\dot a} \lambda_2^{\dot b} &=& {1 \over 8} (\lambda_1^{\dot c}
\lambda_2^{\dot c}) \, \delta_{\dot a \dot b} +
{1 \over 16} (\lambda_1^{\dot c}
\bar \gamma^{IJ}_{\dot c \dot d} \lambda_2^{\dot d} ) \,
\bar \gamma^{IJ}_{\dot a \dot b} +
{1 \over 384} ( \lambda_1^{\dot c }
\bar \gamma^{IJKL}_{\dot c \dot d} \lambda_2^{\dot d} ) \,
\bar \gamma^{IJKL}_{\dot a \dot b}, \label{fiertztwo} \\
\lambda_1^a \lambda_2^{\dot b} &=& {1 \over 8} (\lambda_1^c
\gamma^{I}_{c\dot d} \lambda_2^{\dot d}) \gb_{\dot b a}^{I} -
{1 \over 48}(\lambda_1^c \gamma^{IJK}_{c\dot d} \lambda_2^{\dot d})
\gb_{\dot b a}^{{IJK}}.
\label{fiertzthree}
\eea
These identities hold both in the real and the hermitian
representation described above. These can be proved by using various
representation independent relations summarized below:
\bea
{\rm Tr}(\g^I \gb^J) &=& 8 \delta_{IJ}~, \cr
{\rm Tr}(\g^I \bar \g^{JKL}) &=& 0 ~,  \cr
{\rm Tr}(\g^{IJK} \bar \g^{MNP}) &=& -8 \Del_{(IJK),(MNP)}~, \cr
{\rm Tr} (\gamma^{IJKL}) &=& 0, \cr
{\rm Tr} (\gamma^{IJ} \gamma^{MN}) &=& -8 \Del_{(IJ),(MN)}~, \cr
{\rm Tr} (\gamma^{IJ} \gamma^{MNPQ}) &=& 0, \cr
{\rm Tr} (\gamma^{IJKL} \gamma^{MNPQ}) &=& 8 \Del_{(IJKL),(MNPQ)} + 8
\epsilon^{IJKLMNPQ}~, \cr
\g^1\gb^2\cdots \gb^8 &=& - \gb^1 \g^2 \cdots \g^8 = \one_8 ~,
\label{traces}
\eea
where $\Del_{(I_1,\cdots I_n), (J_1,\cdots J_n)}$ is defined in
eq.(\ref{Delta}). The trace formulas also hold with $\g$ and $\gb$
interchanged.

\section{Results of the Integrals}
\label{a:integrals}
To compute the one-point functions in eqs.(\ref{disk0}) and
(\ref{disk1}, \ref{disk2}, \ref{disk3}, \ref{disk4}) one
needs to evaluate certain integrals where various correlation
functions on the full plane appear in the integrands. These
correlation functions can be written in terms of various functions
that have been computed in appendix \ref{a:correlators}. Below we list
the expected results for all the required integrals containing some of
these functions in the integrands. These results have been tested for various
values of the positive integers $m,n,p,q$ using Mathematica.
\bea
\oint_i {dz \over 2\pi i}
\oint_{-i} {dw \over 2\pi i} {\cal J}_{m,n}(z, w)
\tilde F^{(1,1)}(z,w,i,-i) &=& 8 \delta_{mn} ~, \label{int1} \\
\oint_i {dz \over 2\pi i}
\oint_{-i} {dw \over 2\pi i} {\cal J}_{m,n}(z, w)
\tilde F^{(3,3)}(z,w,i,-i) &=& 8 \delta_{mn} ~, \label{int2}\\
\oint_i {dz_1 dw_1\over (2\pi i)^2}
\oint_{-i} {dz_2 dw_2 \over (2\pi i)^2} {\cal J}_{m,n,p,q}(z_1,w_1,z_2,w_2)
F^{(1,1)}(z_1,w_1,z_2,w_2,i,-i) &=& 8(\delta_{mp}\delta_{nq}-
\delta_{mq}\delta_{np})~, \cr && \label{int3} \\
\oint_i {dz_1 dw_1\over (2\pi i)^2}
\oint_{-i} {dz_2 dw_2 \over (2\pi i)^2} {\cal J}_{m,n,p,q}(z_1,w_1,z_2,w_2)
F^{(3,3)}_1(z_1,w_1,z_2,w_2,i,-i) &=& 8(\delta_{mp}\delta_{nq}+
\delta_{mq}\delta_{np})~, \cr && \label{int4} \\
\oint_i {dz_1 dw_1\over (2\pi i)^2}
\oint_{-i} {dz_2 dw_2 \over (2\pi i)^2} {\cal J}_{m,n,p,q}(z_1,w_1,z_2,w_2)
F^{(3,3)}_2(z_1,w_1,z_2,w_2,i,-i) &=& 0 ~, \label{int5} \\
\oint_i {dz_1 dw_1\over (2\pi i)^2}
\oint_{-i} {dz_2 dw_2 \over (2\pi i)^2} {\cal J}_{m,n,p,q}(z_1,w_1,z_2,w_2)
F^{(3,3)}_2(z_1,w_1,z_2,w_2,-i,i) &=& 0 ~, \label{int6} \\
\oint_i {dz_1 dw_1\over (2\pi i)^2}
\oint_{-i} {dz_2 dw_2 \over (2\pi i)^2} {\cal J}_{m,n,p,q}(z_1,w_1,z_2,w_2)
F^{(5,5)}_1(z_1,w_1,z_2,w_2,i,-i) &=& 8(\delta_{mp}\delta_{nq}-
\delta_{mq}\delta_{np})~, \cr && \label{int7} \\
\oint_i {dz_1 dw_1\over (2\pi i)^2}
\oint_{-i} {dz_2 dw_2 \over (2\pi i)^2} {\cal J}_{m,n,p,q}(z_1,w_1,z_2,w_2)
F^{(5,5)}_2(z_1,w_1,z_2,w_2,i,-i) &=& 0 ~. \label{int8} \\
\oint_i {dz_1 dw_1\over (2\pi i)^2}
\oint_{-i} {dz_2 dw_2 \over (2\pi i)^2} {\cal J}_{m,n,p,q}(z_1,w_1,z_2,w_2)
F^{(5,5)}_2(z_1,w_1,z_2,w_2,-i,i) &=& 0 ~. \label{int9}
\eea

\end{document}